\documentstyle[12pt]{article}

 \makeatletter \textheight 24cm
\newcommand{\bd}{\begin{document}}
\newcommand{\ed}{\end{document}}
\newcommand{\bc}{\begin{center}}
\newcommand{\bt}{\begin{tabbing}}
\newcommand{\ec}{\end{center}}
\newcommand{\et}{\end{tabbing}}
\newcommand{\be}{\begin{eqnarray}}
\newcommand{\ee}{\end{eqnarray}}
\newcommand{\nn}{\nonumber}
\newcommand{\eqn}{\global\def\theequation}
\def\figcap{\section*{Figure Captions\markboth
     {FIGURECAPTIONS}{FIGURECAPTIONS}}\list
     {Figure \arabic{enumi}:\hfill}{\settowidth\labelwidth{Figure 999:}
     \leftmargin\labelwidth
     \advance\leftmargin\labelsep\usecounter{enumi}}}
\let\endfigcap\endlist \relax
\def\reflist{\section*{References\markboth
     {REFLIST}{REFLIST}}\list
     {[\arabic{enumi}]\hfill}{\settowidth\labelwidth{[999]}
     \leftmargin\labelwidth
     \advance\leftmargin\labelsep\usecounter{enumi}}}
\let\endreflist\endlist \relax
\input{tcilatex}

\begin{document}

\title{Axial-vector Form Factors for $K_{l2\gamma }$ and $\pi _{l2\gamma }$\ at $%
O(p^{6}) $ in Chiral Perturbation Theory}
\author{C.Q. Geng$^{1,2}$, I-Lin Ho$^{1}$, and T.H. Wu$^{1}$\footnote{Deceased} \\
\\
$^{1}$Department of Physics, National Tsing Hua University\\
Hsinchu, Taiwan\\
$^{2}$ Theory Group, TRIUMF\\
Vancouver, B.C. V3T 2A3, Canada} \maketitle

\begin{abstract}
We present two-loop calculations on the axial-vector form factors
$F_A$ of semileptonic radiative kaon and pion decays in chiral
perturbation theory. The relevant dimension-6 terms of the
lagrangian are evaluated from the resonance contribution and the
results of the irreducible two-loop graphs of the sunset topology
are given in detail. We also explicitly show that the divergent
parts in $F_A$ are cancelled exactly as required.
\end{abstract}

\section{Introduction}

Chiral perturbation theory (ChPT) \cite{GL1,GL2} has established
itself as a powerful effective theory of low energy interactions.
While it works for the strong interaction, it also includes the
electroweak one whose dynamics can be completely fixed by
introducing the corresponding gauge bosons through the usual
covariant derivative. Since the external momenta and quark masses
are the expansion parameters for the generating function in ChPT
\cite {Pi,BKM,Ec}, they need to be small compared to the physical
scale of the chiral symmetry breaking, i.e. about 1 GeV.
Therefore, one expects that the semileptonic radiative kaon (pion)
decays of $K^+\to l^+\nu_l\gamma$ ($\pi^+\to l^+\nu_l\gamma$) can
be well described in ChPT \cite{Handbook,BEG}.
It is known that these radiative decays \cite{Bo,Po,E787,E246}
could provide us with information on new physics \cite{GL,CGL} by
searching for the lepton polarization effects, which depend on the
vector and axial-vector form factors $F_{V,A}$ of the structure
dependent parts.

In this paper we deal mainly with the $SU(3)\otimes SU(3)$ chiral
symmetry. We will present two-loop calculations on the
axial-vector form factors in $K^+\to l^+\nu_l\gamma$ ($%
K_{l2\gamma })$ and $\pi^+\to l^+\nu_l\gamma$ ($\pi _{l2\gamma })$
by virtue of  the recent progresses in the $p^{6}$-Lagrangian
\cite {BCE1,FS,EFS} and the massive two-loop integrals
\cite{GY,PT}.
 Some remarks related to the form factors in $K_{l2\gamma }$ and $\pi _{l2\gamma
}$ follow:

\begin{itemize}
\item  The usual one-loop ChPT for the timelike form factor does
not satisfy the unitarity or the final-state theorem and makes
poor approximations \cite {Tr1,DHT,Tr2}. Substantial corrections
are expected at a higher order, especially at the two-loop level.

\item  The form factors in the decays receive
the first non-vanishing contributions at $O(p^{4})$ but the first
sizable ones, as well as the estimates of the accuracy, arise at
$O(p^{6})$.

\item  The question of convergence in ChPT needs to be clearly
addressed \cite{BCE2}.

\item In ChPT, the $O(p^6)$ contributions to the vector form
factors $F_V$ in the decays have been studied in Ref. \cite{ABBC},
but those to the axial-vector ones $F_A$ have been done only for
$\pi\to l\nu_l\gamma$ based on the $SU(2)\times SU(2)$ symmetry
\cite{BT}.

\end{itemize}

In this paper, by using the relevant dimension-6 terms of the
lagrangian \cite{PT,DHT} from the resonance contribution
\cite{ABBC,BT,KN}, we perform a detailed calculation for the
irreducible two-loop graphs of the sunset topology \cite{Tr2},
which give the dominant contributions to $F_A$ in both $\pi$ and
$K$ decays at $O(p^6)$. For completeness, we will also evaluate
$F_V$ and compare our results with those in Ref. \cite{ABBC}.

 The
paper is organized as follows. In Sec. 2, we give the matrix
elements for the
decays. 
We review the lagrangians of ChPT to  $O(p^{6})$ in Sec. 3. In
Sec. 4, we display the two-loop calculations for both vector and
axial-vector form factors with the detailed formulas placed in
 Appendices A, B and C. In Sec. 5, we show the analytical results. We
present our numerical values and conclusions in Sec. 6.

\section{The matrix elements}


We consider the decay of $P(p)\rightarrow l^{+}(l^{\prime
})\nu_{l}(s^{\prime })\gamma (k)$ with $P=K^{+}$ or $\pi ^{+}$,
where $\gamma $ is a real photon with $k^{2}=0$. The matrix
element $M$ for the decay \cite{Handbook,DBH1,DBH2} can be written
as
\begin{equation}
M_{P\to l^{+}\nu_{l}\gamma }=-i\frac{eG_{F}}{\sqrt{2}}
\theta_PM_{\mu \nu }(p,k)\epsilon ^{\mu \ast
}(k)\overline{u}(s^{\prime })\gamma ^{\nu }(1-\gamma
_{5})v(l^{\prime })\,,  \label{M1}
\end{equation}
where $\epsilon ^{\mu }$ is the photon polarization and $\theta_{K^+(
\pi^+)}=\cos\theta\, (\sin\theta)$ with $\theta$ being the Cabibbo angle. In
Eq. (\ref{M1}), the hadronic part of the quantity $M_{\mu \nu }$ is given by
\begin{eqnarray}
M_{\mu \nu }(p,k) &=&\int d^{4}xe^{iq\cdot x}\left\langle 0\left| T(J_{\mu
}^{em}(x)J_{\nu }^{wk}(0))\right| P(p)\right\rangle\,,
\end{eqnarray}
which has the general structure

\begin{eqnarray}
M_{\mu \nu }(p,k)&=&-\sqrt{2}F_P\frac{(p-k)_{\nu }}{(p-k)^{2}-M_{P}^{2}}%
\left\langle P(p-k)\left| J_{\mu }^{em}\right| P(p)\right\rangle +\sqrt{2}%
F_Pg_{\mu \nu }  \nonumber \\
&&-F_{A}[(p-k)_{\mu }k_{\nu }-g_{\mu \nu }k\cdot (p-k)]-r_{A}(k_{\mu }k_{\nu
}-g_{\mu \nu }k^{2})  \nonumber \\
&&+iF_{V}\varepsilon _{\mu \nu \alpha \beta }k^{\alpha }p^{\beta
}\, \label{MFF}
\end{eqnarray}
where the first line represents the Born diagram, in which the
photon couples to hadrons through the known $KK\gamma\,\, (\pi \pi
\gamma )$ coupling, with $F_P$ being the $P$ meson decay constant,
and the subsequent lines correspond to axial-vector and vector
portions of the weak currents with $F_{V(A)}$ being the vector
(axial-vector) form factor. In Eq. (\ref {MFF}), $r_{A}$ is
non-zero only for those processes with virtual photons, such as
$P^-\rightarrow l^{+}l^{+}l^{-}\overline{\nu }_{l}$. In terms of
the form factors $F_{A}$ and $F_{V}$, we can write the vector and
axial-vector parts in Eqs. (\ref{M1}) and (\ref{MFF}) as

\begin{eqnarray}
M_{V}(P &\rightarrow &l^{+}\nu _{l}\gamma )=\frac{eG_{F}\theta _{P}}{\sqrt{2}%
}F_{V}\varepsilon _{\mu \nu \alpha \beta }\epsilon ^{\mu \ast }l^{\nu
}k^{\alpha }p^{\beta }\,,  \label{FV} \\
M_{A}(P &\rightarrow &l^{+}\nu _{l}\gamma )=i\frac{eG_{F}\theta _{P}}{\sqrt{2%
}}F_{A}l^{\nu }[k_{\nu }(p\cdot \epsilon )-g_{\mu \nu }(p\cdot k)\epsilon
^{\mu \ast }]\,,  \label{FA}
\end{eqnarray}
respectively, where
\begin{equation}
l^{\nu }=\overline{u}(s^{\prime })\gamma ^{\nu }(1-\gamma _{5})v(l^{\prime
})\,.
\end{equation}
Both $F_{A}$ and $F_{V}$ 
are real functions for $q^{2}\equiv (p-k)^{2}$ below the physical
threshold, which is the region of interest here, based on
time-reversal invariance, and they are analytic functions of
$q^{2}$ with cuts on the positive real axis. One of the reasons to
perform the present calculation is that the $q^{2}$ dependence of
the form factors starts at $O(p^{6})$.

\section{The Lagrangians of chiral perturbation theory}

In the usual formulation of ChPT \cite{GL1,GL2,PS} with the chiral symmetry $%
SU(3)_{L}\times SU(3)_{R}$, the pseudoscalar fields are collected in a
unitary $3\times 3$ matrix

\begin{equation}
U(x)=exp\left(i\frac{\Phi (x)}{F}\right)\,,
\end{equation}
where F absorbs the dimensional dependence of the fields and, in the chiral
limit, is equal to the pion decay constant, $F_{\pi}=92.4$ MeV. The $\Phi $
is given by the $3\times 3$ matrix

\begin{equation}
\Phi =\lambda _{a}\varphi _{a}=\left(
\begin{array}{ccc}
\pi ^{0}+\frac{\eta }{\sqrt{3}} & \sqrt{2}\pi ^{+} & \sqrt{2}K^{+} \\
\sqrt{2}\pi ^{-} & -\pi ^{0}+\frac{\eta }{\sqrt{3}} & \sqrt{2}K^{0} \\
\sqrt{2}K^{-} & \sqrt{2}\overline{K}^{0} & \frac{-2\eta
}{\sqrt{3}}
\end{array}\,,
\right)
\end{equation}
where $\lambda _{a}\ (a=1,2,\cdots, 8)$ are the Gell-Mann matrices.

An explicit breaking of the chiral symmetry is introduced via the
mass matrix

\begin{equation}
\chi =\left(
\begin{array}{ccc}
m_{\pi }^{2} & 0 & 0 \\
0 & m_{\pi }^{2} & 0 \\
0 & 0 & 2m_{K}^{2}-m_{\pi }^{2}
\end{array}\,,
\right)  \label{chi}
\end{equation}
where $m_{\pi (K)}$ is the unrenormalized $\pi\ (K)$ mass.  We
note that the mass of $\eta$ to this order is given by the
Gell-Mann-Okubo relation
\begin{equation}
m_{\eta }^{2}=\frac{4}{3}m_{K}^{2}-\frac{1}{3}m_{\pi }^{2}\,. %
\label{GO}
\end{equation}
The mass term in Eq. (\ref{chi}) is related to the quark masses by $\chi =$ $%
const\cdot diag(m_{u},m_{d},m_{s})$ with $m_{u}=m_{d}$. To calculate the
form factors, we have to include the interaction with external boson fields.
As previously stated, the electroweak gauge fields $A_{\mu }$ and $W_{\mu }$
are introduced via the covariant derivative

\begin{eqnarray}
D_{\mu }U &=&\partial _{\mu }U+iU\ell _{\mu }-ir_{\mu }U\,,  \nonumber \\
\ell _{\mu } &=&eA_{\mu }Q-\frac{g}{\sqrt{2}}\left(
\begin{array}{ccc}
0 & \cos \theta W_{\mu }^{+} & \sin \theta W_{\mu }^{+} \\
\cos \theta W_{\mu }^{-} & 0 & 0 \\
\sin \theta W_{\mu }^{-} & 0 & 0
\end{array}
\right) \,,  \nonumber \\
r_{\mu }&=&eA_{\mu }Q\,,
\end{eqnarray}
where $Q$ is the quark charge matrix in units of
$e=g\sin\theta_W$, with $\theta
_{W}$ standing for the Weinberg angles and $G_{F}/\sqrt{2}=g^{2}/(8M_{W}^2)$%
. The final lepton pair (for clearity we will always refer to
$l^{+}\nu_l $ coming from $W^{+}$) appears in the leptonic current
when substituting
\begin{equation}
W_{\mu }\rightarrow \frac{g}{2\sqrt{2}M_{W}^{2}}l_{\mu }=\frac{g}{2\sqrt{2}%
M_{W}^{2}}\overline{u}(s)\gamma _{\mu }(1-\gamma _{5})v(l)\,.
\end{equation}

The Lagrangians of ChPT contain both normal (or non-anomalous) and
anomalous parts. Since the form factor $F_{V}$ is related by an
isospin rotation to the amplitude for $\pi ^{0}\rightarrow \gamma
\gamma $, it can be absolutely predicted from the axial anomaly.
For this reason, we must also include the effect of the axial
anomaly. At the two lowest orders, the full non-anomalous
Lagrangian is given by \cite{GL1,GL2}

\begin{eqnarray}
{\cal L}_{n}^{(2)} &=&\frac{F^{2}}{4}Tr(D_{\mu }UD^{\mu }U^{\dagger })+\frac{F^{2}}{%
4}Tr(\chi U^{\dagger }+U\chi ^{\dagger })\,,
  \label{L2} \\
  {\cal
L}_{n}^{(4)} &=&L _{1}\left[ Tr(D_{\mu }UD^{\mu }U^{\dagger
})\right] ^{2}+L _{2}Tr(D_{\mu }UD_{\nu }U^{\dagger })Tr(D^{\mu
}UD^{\nu }U^{\dagger })  \nonumber \\ &&+L _{3}Tr(D_{\mu }UD^{\mu
}U^{\dagger }D_{\nu }UD^{\nu }U^{\dagger }) \nonumber \\ &&+L
_{4}Tr(D_{\mu }UD^{\mu }U^{\dagger })Tr(\chi U^{\dagger }+U\chi
^{\dagger })  \nonumber \\ &&+L _{5}Tr(D_{\mu }UD^{\mu }U^{\dagger
}(\chi U^{\dagger }+U\chi
^{\dagger }))+L _{6}\left[ Tr(\chi U^{\dagger }+U\chi ^{\dagger })%
\right] ^{2}  \nonumber \\ &&+L _{7}\left[ Tr(\chi ^{\dagger
}U-U^{\dagger }\chi )\right] ^{2}+L _{8}Tr(\chi U^{\dagger }\chi
U^{\dagger }+U\chi ^{\dagger }U\chi ^{\dagger })  \nonumber \\
&&+iL _{9}Tr(L_{\mu \nu }D^{\mu }UD^{\nu }U^{\dagger }+R_{\mu \nu
}D^{\mu }U^{\dagger }D^{\nu }U)+L _{10}Tr(L_{\mu \nu }UR_{\mu \nu
}U^{\dagger })\,,  \label{L4}
\end{eqnarray}
where $L_{\mu \nu }$ and $R_{\mu \nu }$ are the field-strength tensors of
external sources, defined by
\begin{eqnarray}
L_{\mu \nu }&=&\partial _{\mu }\ell _{\nu }-\partial _{\nu }\ell _{\mu }-i
\left[ \ell _{\mu },\ell _{\nu }\right] \,,  \nonumber \\
R_{\mu\nu }&=&\partial _{\mu }r_{\nu }-\partial _{\nu }r_{\mu }-i\left[
r_{\mu },r_{\nu }\right]\,,
\end{eqnarray}
and $\{L _{i}\}$ 
are unrenormalized coupling constants.
At $O(p^{6})$, the non-anomalous Chiral Lagrangian
contains 90 independent terms plus four contact terms for $SU(3)$
\cite{PT}. The terms relevant to $K_{l2\gamma }$ $(\pi _{l2\gamma
})$\ decays are found to be

\begin{eqnarray}
{\cal L}_{n}^{(6)} &=&y_{17}\left\langle \chi _{+}h_{\mu \nu
}h^{\mu \nu }\right\rangle +y_{18}\left\langle \chi
_{+}\right\rangle \left\langle h_{\mu \nu }h^{\mu \nu
}\right\rangle +y_{81}\left\langle \chi _{+}f_{+\mu \nu
}f_{+}^{\mu \nu }\right\rangle  \nonumber \\ &&+y_{82}\left\langle
\chi _{+}\right\rangle \left\langle f_{+\mu \nu }f_{+}^{\mu \nu
}\right\rangle +iy_{83}\left\langle f_{+\mu \nu }\left\{ \chi
_{+},u^{\mu }u^{\nu }\right\} \right\rangle +iy_{84}\left\langle
\chi _{+}\right\rangle \left\langle f_{+\mu \nu }u^{\mu }u^{\nu
}\right\rangle \nonumber \\ &&+iy_{85}\left\langle f_{+\mu \nu
}u^{\mu }\chi _{+}u^{\nu }\right\rangle
+iy_{100}\left\langle f_{+\mu \nu }\left[ f_{-}^{\nu \rho },h_{\rho }^{\mu }%
\right] \right\rangle +y_{102}\left\langle \chi _{+}f_{-\mu \nu }f_{-}^{\mu
\nu }\right\rangle  \nonumber \\
&&+y_{103}\left\langle \chi _{+}\right\rangle \left\langle f_{-\mu \nu
}f_{-}^{\mu \nu }\right\rangle +y_{104}\left\langle f_{+\mu \nu }\left[
f_{-}^{\mu \nu },\chi _{-}\right] \right\rangle +y_{109}\left\langle \nabla
_{\rho }f_{-\mu \nu }\nabla ^{\rho }f_{-}^{\mu \nu }\right\rangle  \nonumber
\\
&&+iy_{110}\left\langle \nabla _{\rho }f_{+\mu \nu }\left[ h^{\mu \rho
},u^{\nu }\right] \right\rangle +...\,,  \label{L6}
\end{eqnarray}
where
\begin{eqnarray}
u_{\mu } &=&i\left\{ U^{\dagger }(\partial _{\mu }-ir_{\mu })U-U(\partial
_{\mu }-i\ell _{\mu })U^{\dagger }\right\} \,,  \nonumber \\
\chi _{\pm }&=& U^{\dagger }\chi U^{\dagger }+U\chi ^{\dagger }U \,,
\nonumber \\
f_{\pm }^{\mu \nu } &=&UL^{\mu \nu }U^{\dagger }\pm U^{\dagger }R^{\mu \nu
}U\,,  \nonumber \\
h_{\mu \nu }&=&\nabla _{\mu }u_{\nu }+\nabla _{\nu }u_{\mu }\,,  \nonumber \\
\chi _{\pm \mu } &=&U^{\dagger }D_{\mu }\chi U^{\dagger }\pm UD_{\mu }\chi
^{\dagger }U=\nabla _{\mu }\chi _{\pm }-\frac{i}{2}\left\{ \chi _{\mp
},u_{\mu }\right\}\,.
\end{eqnarray}
The covariant derivative $\nabla _{\mu }X=\partial _{\mu }X+\left[ \Gamma
_{\mu },X\right] $ is defined in terms of the chiral condition

\begin{equation}
\Gamma _{\mu }=\frac{1}{2}\left\{ U^{\dagger }(\partial _{\mu }-ir_{\mu
})U-U(\partial _{\mu }-i\ell _{\mu })U^{\dagger }\right\}\,.
\end{equation}

The anomalous part begins at $O(p^{4})$ with the Wess-Zumino
(WZ) term ${\cal L}_{WZ}$ %
\cite{St} containing pieces with zero, one and two gauge boson
fields. The terms with one and two gauge bosons
as well as the anomalous $p^{6}$-Lagiangian \cite{DHT,St} are
given by

\begin{eqnarray}
{\cal L}_{WZ,1}^{(4)} &=&-\frac{1}{16\pi ^{2}}\varepsilon ^{\mu
\nu \alpha \beta }tr(U\partial _{\mu }U^{\dagger }\partial _{\nu
}U\partial _{\alpha }U^{\dagger }\ell _{\beta }-U^{\dagger
}\partial _{\mu }U\partial _{\nu }U^{\dagger }\partial _{\alpha
}Ur_{\beta })\,,
 \\ {\cal L}_{WZ,2}^{(4)} &=&-\frac{i}{16\pi
^{2}}\varepsilon ^{\mu \nu \alpha \beta }tr(\partial _{\mu
}U^{\dagger }\partial _{\nu }\ell _{\alpha }Ur_{\beta }-\partial
_{\mu }U\partial _{\nu }r_{\alpha }U^{\dagger }\ell _{\beta })
\nonumber \\ &&+U\partial _{\mu }U^{\dagger }(\ell _{\nu }\partial
_{\alpha }\ell _{\beta }+\partial _{\nu }\ell _{\alpha }\ell
_{\beta })\,,
 \\ {\cal L}_{a}^{(6)} &=&iC_{7}\varepsilon ^{\mu \nu
\alpha \beta }\left\langle \chi _{-}f_{+\mu \nu }f_{+\alpha \beta
}\right\rangle +iC_{11}\varepsilon ^{\mu \nu \alpha \beta
}\left\langle \chi _{-}[f_{+\mu \nu },f_{-\alpha \beta
}]\right\rangle  \nonumber \\ &&+C_{22}\varepsilon ^{\mu \nu
\alpha \beta }\left\langle u_{\mu }\{\nabla _{\gamma }f_{+\gamma
\nu },f_{+\alpha \beta }\}\right\rangle +\cdots\,.
\end{eqnarray}
 From
 the above expressions, the Feynman rules can be derived
by expanding $U=exp(i\Phi/F)$ everywhere in ${\cal L}={\cal
L}^{(2)}+{\cal L}^{(4)}+{\cal L}^{(6)}$ and identifying the
relevant vertex monomials.
In the next section we display the main result of the two-loop
calculations for the form factors.

\section{The Form Factors}


To $O(p^{6})$, the finite matrix elements in ChPT are obtained by
multiplying the unrenormalized Feynman diagrams obtained from $%
{\cal L}={\cal L}^{(2)}+{\cal L}^{(4)}+{\cal L}^{(6)}$ with a
factor $\sqrt{Z}$ per external meson, where $Z$ is the wave
function renormalization constant. To get these results, we start
to calculate the mass and wave function renormalizations as well
as that of the pion decay constant.

Since the form factors at $O(p^{4})$ are related to the finite counterterm
contribution, we only need the wave function renormalization, $m_{K,\pi
}^{2} $ and $F_{K,\pi }$ to $O(p^{4})$ in our calculations. Explicitly, we
have

\begin{eqnarray}
Z_{\pi }^{-1} &=&1-\frac{1}{3F^{2}}\left[I(m_{K}^{2})+2I(m_{\pi
}^{2})-24(2m_{K}^{2}+m_{\pi }^{2})L _{4}-24m_{\pi }^{2}L_{
5}\right]\,,%
 \\ %
  Z_{K}^{-1} &=&1-\frac{1}{4F^{2}}\left[I(m_{\eta
}^{2})+2I(m_{K}^{2})+I(m_{\pi }^{2})-32(2m_{K}^{2}+m_{\pi
}^{2})L _{4}-32m_{K}^{2}L_{5}\right]\,,%
 \\ %
  \delta m_{\pi }
&=&\frac{1}{6F^{2}}\left[-m_{\pi }^{2}I(m_{\eta }^{2})+3m_{\pi
}^{2}I(m_{\pi }^{2})-48m_{\pi }^{2}(2m_{K}^{2}+m_{\pi }^{2})
L _{4}-48L _{5}m_{\pi }^{4} \right.%
\nonumber \\ %
&&\left.+96L _{6}m_{\pi }^{2}(2m_{K}^{2}+m_{\pi
}^{2})+96L _{8}m_{\pi }^{4}\right]\,, %
\\ %
 \delta m_{K}
&=&\frac{1}{12F^{2}}\left[4m_{K}^{2}I(m_{\eta }^{2})-96L
_{4}m_{K}^{2}(2m_{K}^{2}+m_{\pi }^{2})-96L _{5}m_{K}^{4}+192L _{8}m_{K}^{4}\right.%
  \nonumber \\ %
&&\left.+192L _{6}m_{K}^{2}(2m_{K}^{2}+m_{\pi }^{2})\right]\,,
\\ F_{\pi }
&=&F\left\{1+\frac{1}{F^{2}}\left[-\frac{1}{2}I(m_{K}^{2})-I(m_{\pi
}^{2})+4(2m_{K}^{2}+m_{\pi }^{2})L _{4}+4m_{\pi }^{2}L
_{5}\right]\right\}\,,
\\
F_{K} &=&F\left\{1+\frac{1}{F^{2}}\left[-\frac{3}{8}I(m_{\eta }^{2})-\frac{3}{4}%
I(m_{K}^{2})-\frac{3}{8}I(m_{\pi }^{2})+4(2m_{K}^{2}+m_{\pi }^{2})L _{4}%
\right.\right.%
 \nonumber \\ %
  &&\left.\left.+4m_{K}^{2}L _{5}\right]\right\}\,,
\end{eqnarray}
where
\begin{eqnarray}
I(m^{2})\equiv \mu ^{4-D}\int \frac{d^{D}q}{(2\pi )^{D}}\frac{i}{q^{2}-m^{2}}%
=-\frac{m^{2}}{16\pi ^{2}}\left[\frac{1}{\varepsilon }+1+\ln (4\pi )-\gamma
-\ln \left(%
\frac{m^{2}}{\mu ^{2}}\right)\right]
\end{eqnarray}
is the standard tadpole integral. It should be noted that the
renormalization constants $\delta m(\equiv m_{phys}-m)$ and
$\delta F_{K,\pi }(\equiv F_{K,\pi }-F)$ defined above are finite.
The divergences and scale dependences of the loop integrals are
cancelled by similar factors for $L _{i}$ in ${\cal L}^{(4)}$.

\subsection{The vector form factors $F_{V}$}

Using the chiral Lagrangians mentioned above, one immediately
obtains the tree-level contribution for the anomalous parts of our
processes. For the semileptonic radiative $K$ decays, we have

\begin{equation}
F_{V,K,tree}=\frac{1}{4\sqrt{2}\pi ^{2}F}\left[1-\frac{256}{3}\pi
^{2}m_{K}^{2}C_{7}^{W}+256\pi ^{2}(m_{K}^{2}-m_{\pi }^{2})C_{11}^{W}+\frac{64%
}{3}\pi ^{2}(q^{2}+k^{2})C_{22}^{W}\right]\,.
\label{n29}
\end{equation}
 Loop corrections to the above tree-level
contribution proceed through diagrams involving at least one
vertex given by the WZ lagrangian.
\begin{figure}[h]
\begin{center}
\vspace{2cm}
\includegraphics{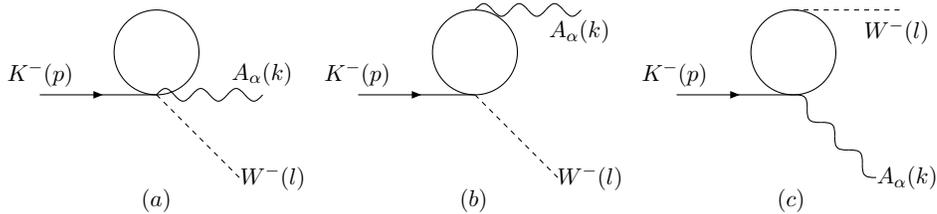}
\end{center}
\caption{One-loop diagrams that contribute to $F_V$ in $P_{\ell
2\gamma } $} \label{fig1}
\end{figure}
As shown in Figure \ref{fig1} with $P=K$, due to the initial
order, most one-loop diagrams can contribute to $F_V$ to
$O(p^{6})$ constructed with  one vertex coming from the WZ
Lagrangian and another one from the lowest order chiral
lagrangian. Performing the calculation in the three-flavor case,
we
get
\begin{eqnarray}
F_{V,K,loop(a)}=\frac{1}{4\sqrt{2}\pi ^{2}F}\cdot \frac{1}{16\pi ^{2}(\sqrt{2%
}F)^{2}}\left[\lambda (m_{\eta }^{2}+2m_{K}^{2}+3m_{\pi }^{2})
~~~~~~~~~~~~~~~~~~~~~~~~~~~~~\right.  \nonumber \\
\left.~~~~~~~~~~~~~~~~~~~~~~~ -m_{\eta }^{2}\ln \left(\frac{m_{\eta }^{2}}{\mu ^{2}%
}\right)-2m_{K}^{2}\ln \left(\frac{m_{K}^{2}}{\mu
^{2}}\right)-3m_{\pi }^{2}\ln\left( \frac{m_{\pi }^{2}}{\mu
^{2}}\right)\right]\,, \nonumber \\
F_{V,K,loop(b)}=\frac{1}{4\sqrt{2}\pi ^{2}F}\cdot \frac{1}{16\pi ^{2}(\sqrt{2%
}F)^{2}}\left[(-4m_{\pi }^{2}+\frac{2}{3}k^{2})\lambda
~~~~~~~~~~~~~~~~~~~~~~~~~~~~~~~~~~~ \right.  \nonumber \\
\left. ~~~~~~~~~~~~~~~~~~~~~~~~~ +4\int (m_{\pi }^{2}-x(1-x)k^{2})\ln \left(\frac{%
m_{\pi }^{2}-x(1-x)k^{2}}{\mu ^{2}}\right)\right]\,,  \nonumber \\
F_{V,K,loop(c)}=\frac{1}{4\sqrt{2}\pi ^{2}F}\cdot \frac{1}{16\pi ^{2}(\sqrt{2%
}F)^{2}}\left[(-m_{\pi }^{2}-m_{\eta }^{2}-2m_{K}^{2}+\frac{2}{3}%
q^{2})\lambda ~~~~~~~~~~~~~~~~~~~~ \right.
\nonumber \\ +2\int
(xm_{\eta }^{2}+(1-x)m_{K}^{2}-x(1-x)q^{2})\ln
\left(\frac{xm_{\eta }^{2}+(1-x)m_{K}^{2}-x(1-x)q^{2}}{\mu
^{2}}\right)dx ~
\nonumber \\ \left. +2\int (xm_{\pi
}^{2}+(1-x)m_{K}^{2}-x(1-x)q^{2})\ln \left(\frac{xm_{\pi
}^{2}+(1-x)m_{K}^{2}-x(1-x)q^{2}}{\mu ^{2}}\right)dx \right]\,,
\end{eqnarray}
from the three loops in Figure \ref{fig1}, respectively, where
\begin{eqnarray}
\lambda &=&\frac{1}{\varepsilon }+1+\ln (4\pi )-\gamma \,.
\end{eqnarray}
Putting all of the contributions together as well as the usual
renormalizations of the pseudoscalar wave-functions and decay
constants, we obtain the following results:

\begin{eqnarray}
F_{V,K} =\frac{1}{4\sqrt{2}\pi ^{2}F_{K}}\left\{1-\frac{256}{3}\pi
^{2}m_{K}^{2}C_{7}+256\pi ^{2}(m_{K}^{2}-m_{\pi
}^{2})C_{11}+\frac{64}{3}\pi ^{2}(q^{2}+k^{2})C_{22} \right.
~~~~~~~~~%
\nonumber \\ +\frac{1}{16\pi
^{2}(\sqrt{2}F_{K})^{2}}\left[\frac{5}{3}(m_{K}^{2}-m_{\pi
}^{2})\lambda +\frac{2}{3}(k^{2}+q^{2})\lambda -\frac{3}{2}m_{\eta
}^{2}\ln \left(\frac{m_{\eta }^{2}}{\mu ^{2}}
\right)-3m_{K}^{2}\ln \left(\frac{m_{K}^{2}}{\mu
^{2}}\right)\right.
~~
\nonumber \\ %
-\frac{7}{2}m_{\pi }^{2}
\ln \left(\frac{%
m_{\pi }^{2}}{\mu ^{2}}\right)+4\int \left[m_{\pi
}^{2}-x(1-x)k^{2}\right]\ln \left(\frac{m_{\pi
}^{2}-x(1-x)k^{2}}{\mu ^{2}}\right) ~~~~~~~~~~~~~~~~~~~~~~~~  %
\nonumber \\ %
 +2\int
\left[xm_{\pi }^{2}+(1-x)m_{K}^{2}-x(1-x)q^{2}\right]\ln
\left(\frac{xm_{\pi }^{2}+(1-x)m_{K}^{2}-x(1-x)q^{2}}{\mu
^{2}}\right)%
 ~~~~~~~ \nonumber \\
 \left.\left.+2\int \left[xm_{\eta
}^{2}+(1-x)m_{K}^{2}-x(1-x)q^{2}\right]\ln \left(\frac{xm_{\eta
}^{2}+(1-x)m_{K}^{2}-x(1-x)q^{2}}{\mu
^{2}}\right)\right]\right\}\,. ~~\label{FVK}
\end{eqnarray}
Similarly, for $\pi _{l2\gamma }$ we derive
\begin{eqnarray}
F_{V,\pi } =\frac{1}{4\sqrt{2}\pi ^{2}F_{\pi }}\left\{1-\frac{256}{3}\pi
^{2}m_{\pi }^{2}C_{7}+\frac{64}{3}\pi ^{2}(q^{2}+k^{2})C_{22} +\frac{1}{%
16\pi ^{2}(\sqrt{2}F_{\pi })^{2}}\left[ ~~~~~~~~~~~~\right.\right.  \nonumber
\\
-4m_{K}^{2}\ln \frac{m_{K}^{2}}{\mu ^{2}}-4m_{\pi }^{2}\ln \frac{m_{\pi }^{2}%
}{\mu ^{2}}+4\int (m_{K}^{2}-x(1-x)k^{2})\ln \frac{m_{K}^{2}-x(1-x)k^{2}}{%
\mu ^{2}} ~~  \nonumber \\
\left.\left.+4\int (m_{\pi }^{2}-x(1-x)q^{2})\ln \frac{m_{\pi
}^{2}-x(1-x)q^{2}}{\mu ^{2}}+\frac{2}{3}(k^{2}+q^{2})\lambda \right]%
\right\}\,.~~~~~~~~~~~~~~~  \label{FVP}
\end{eqnarray}
We note that the numerical results of Eqs. (\ref{FVK}) and
(\ref{FVP}) agree with those in Ref. \cite{ABBC} as will be shown
in Sec. 6.
The one-loop corrections contain divergent terms proportional to $\lambda =%
\frac{1}{\varepsilon }+1+\ln (4\pi )-\gamma $ coming from the
dimensional regularization scheme. Obviously, the presence of
these divergent terms requires the introduction of the
corresponding counterterms in the anomalous section of the
Lagrangian at $O(p^{6})$, which were already given in
Refs. %
\cite{DHT,BBC1,BBC2}. Their infinite parts cancel the $\lambda $-terms in Eqs. (%
\ref{FVK}) and (\ref{FVP}) and the coefficients $C_{i}^{W}$\ are
simply
substituted for by the remaining finite parts, the renormalized  coefficients
$C_{i}^{Wr}$%
. The values of these finite contributions from the counterterms
to our processes are not pinned down in ChPT and they have to be
deduced from data fitting \cite{St} or, alternatively, from the
hypothesis of resonance saturation (RS) of the counterterms
\cite{ABBC}. The estimations of using RS are described in Sec. 6.

\subsection{The axial-vector form factors $F_{A}$}

In this subsection we aim at the extraction of the form factor
$F_{A}$, which is the only one that has a contribution
proportional to $g_{\mu \nu }(p\cdot k)$. The presence of $g_{\mu
\nu }$ requires that the axial-vector and vector insertions are in
the same one-particle irreducible subdiagrams. This immediately
removes a large part of the diagrams. Furthermore, the $(p\cdot
k)$ kinematical factor guarantees that it is not part of the
internal Bremsstrahlung contribution. We now discuss the
contributions from the diagrams to $O(p^{6})$.

\subsubsection{Tree-level diagrams}

With the chiral Lagrangians in Eqs. (\ref{L2}), (\ref{L4}) and
(\ref{L6}), one obtains tree-level contributions for the processes
as
follows: 
\begin{eqnarray}
F_{A,\pi ,tree}^{(4)} &=&\frac{4\sqrt{2}}{F}(L _{9}+L _{10})\,,
\\ F_{A,K,tree}^{(4)} &=&\frac{4\sqrt{2}}{F}(L _{9}+L _{10})\,,
\\
F_{A,\pi ,tree}^{(6)} &=&-y_{17}\frac{48\sqrt{2}m_{\pi }^{2}}{F}-y_{18}\frac{%
48\sqrt{2}(2m_{K}^{2}+m_{\pi }^{2})}{F}+y_{81}\frac{16\sqrt{2}m_{\pi }^{2}}{F%
}  \nonumber \\
&&+y_{82}\frac{16\sqrt{2}(2m_{K}^{2}+m_{\pi }^{2})}{F} -y_{83}\frac{16\sqrt{2%
}m_{\pi }^{2}}{F}-y_{84}\frac{8\sqrt{2}(2m_{K}^{2}+m_{\pi }^{2})}{F}
\nonumber \\
&&-y_{85}\frac{8\sqrt{2}m_{\pi }^{2}}{F}+y_{100}\frac{8\sqrt{2}(p_{\pi
}^{2}-p_{\pi }\cdot k)}{F} -y_{102}\frac{16\sqrt{2}m_{\pi }^{2}}{F}
\nonumber \\
&&-y_{103}\frac{16\sqrt{2}(2m_{K}^{2}+m_{\pi }^{2})}{F}+y_{104}\frac{16\sqrt{%
2}m_{\pi }^{2}}{F}+y_{109}\frac{8\sqrt{2}(2p_{\pi }\cdot k-p_{\pi }^{2})}{F}
\nonumber \\
&&-y_{110}\frac{4\sqrt{2}p_{\pi }\cdot k}{F}\,,  \label{FAT6P} \\
F_{A,K,tree}^{(6)} &=&-y_{17}\frac{48\sqrt{2}m_{K}^{2}}{F}-y_{18}\frac{48%
\sqrt{2}(2m_{K}^{2}+m_{\pi }^{2})}{F}+y_{81}\frac{16\sqrt{2}%
(2m_{K}^{2}+m_{\pi }^{2})}{3F}  \nonumber \\
&&+y_{82}\frac{16\sqrt{2}(2m_{K}^{2}+m_{\pi }^{2})}{F} -y_{83}\frac{16\sqrt{2%
}(2m_{K}^{2}+m_{\pi }^{2})}{3F}  \nonumber \\
&&-y_{84}\frac{8\sqrt{2}(2m_{K}^{2}+m_{\pi }^{2})}{F}-y_{85}\frac{8\sqrt{2}%
(4m_{K}^{2}-m_{\pi }^{2})}{3F}  \nonumber \\
&&+y_{100}\frac{8\sqrt{2}(p_{K}^{2}-p_{K}\cdot k)}{F} -y_{102}\frac{16\sqrt{2%
}m_{K}^{2}}{F}-y_{103}\frac{16\sqrt{2}(2m_{K}^{2}+m_{\pi }^{2})}{F}
\nonumber \\
&&+y_{104}\frac{16\sqrt{2}m_{K}^{2}}{F}+y_{109}\frac{8\sqrt{2}(2p_{K}\cdot
k-p_{K}^{2})}{F} -y_{110}\frac{4\sqrt{2}p_{K}\cdot k}{F}\,.  \label{FAT6K}
\end{eqnarray}
We note that for the tree-level contributions in Eqs.
(\ref{FAT6P}) and (\ref{FAT6K}) at $O(p^{6})$ one needs to perform
renormalization with the finite parts, which will be discussed in
Sec. 5.

\subsubsection{One-loop diagrams}

We now consider the one-loop diagrams shown in Figure \ref{fig1}.
Since it contains
only  photon-even-meson vertices in the non-anomalous chiral lagrangian to $%
O(p^{4})$, Figure \ref{fig1}(c) does not contribute to $F_{A}$.
Moreover, one-loop diagrams with an $O(p^{4})$ vertex insertion on
a propagator in the loop never produce the factor $(p\cdot k)$ and
hence do not contribute to $F_{A}$. From Figures \ref{fig1}(a) and
\ref{fig1}(b), we get
\begin{eqnarray}
F_{A,\pi ,loop(a)} &=&-\frac{L _{9}}{3F^{3}}\left[ 14\sqrt{2}%
I(m_{K}^{2})+28\sqrt{2}I(m_{\pi }^{2})\right]   \nonumber \\
&&-\frac{2L _{10}}{3F^{3}}\left[ 10\sqrt{2}I(m_{K}^{2})+20\sqrt{2}%
I(m_{\pi }^{2})\right] \,,  \nonumber \\ F_{A,K,loop(a)}
&=&-\frac{L _{9}}{3F^{3}}\left[ 6\sqrt{2}I(m_{\eta
}^{2})+24\sqrt{2}I(m_{K}^{2})+12\sqrt{2}I(m_{\pi }^{2})\right]
\nonumber \\
&&-\frac{2L _{10}}{3F^{3}}\left[ 3\sqrt{2}I(m_{\eta }^{2})+18\sqrt{2}%
I(m_{K}^{2})+9\sqrt{2}I(m_{\pi }^{2})\right] \,,  \nonumber \\
F_{A,\pi ,loop(b)} &=&\frac{-16\sqrt{2}L _{1}}{F^{3}}I(m_{\pi }^{2})+%
\frac{8\sqrt{2}L _{2}}{F^{3}}I(m_{\pi }^{2})-\frac{8\sqrt{2}L _{3}%
}{F^{3}}\left[ I(m_{\pi }^{2})+\frac{1}{2}I(m_{K}^{2})\right]   \nonumber \\
&&-\frac{2\sqrt{2}L _{9}}{F^{3}}\left[ I(m_{K}^{2})+2I(m_{\pi }^{2})%
\right]\,,   \nonumber \\
F_{A,K,loop(b)} &=&\frac{-16\sqrt{2}L _{1}}{F^{3}}I(m_{K}^{2})+\frac{8%
\sqrt{2}L _{2}}{F^{3}}I(m_{K}^{2})-\frac{8\sqrt{2}L _{3}}{F^{3}}%
\left[ I(m_{K}^{2})+\frac{1}{2}I(m_{\pi }^{2})\right]   \nonumber \\
&&-\frac{2\sqrt{2}L _{9}}{F^{3}}\left[ I(m_{\pi }^{2})+2I(m_{K}^{2})%
\right] \,.
\end{eqnarray}

\subsubsection{Two-loop diagrams}

\begin{figure}[h]
\begin{center}
\vspace{1.3cm} \includegraphics{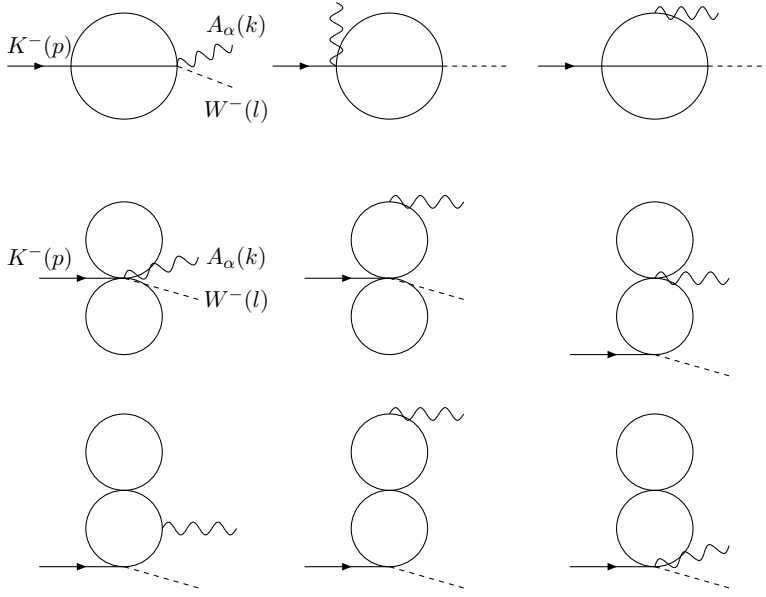}
\end{center}
\par
\vspace{5.7cm} \caption{Two-loop diagrams for the form factors in
$P _{\ell 2\gamma }$. } \label{fig2}
\end{figure}

The two-loop diagrams which may contribute to $F_A$ are shown in
Figure \ref {fig2}. The last six diagrams with nonoverlapping
loops in Figure \ref {fig2}, which can be written as the products
of one-loop integrals and produce no $(p\cdot k)$ factor, do not
contribute to $F_A$. The only possible non-vanishing diagrams are
the first three irreducible ones in Figure \ref {fig2}.

Since there are three different mass scales of $(m_{\pi
},m_{K},m_{\eta })$ with the same order of magnitude in the
$SU(3)$ ChPT, the irreducible integrals can no longer be expressed
by elementary analytical functions. We will quote only the
numerical results in Sec. 5 and Sec. 6.
We now give the detailed calculations for the $g_{\mu\nu}$ terms
of $M_A$ in Eq. (\ref{FA}). It is clear that the first irreducible
diagram in Figure \ref{fig2} does not contribute to $F_A$ since
there is no $(p\cdot k)$ term. The second and third irreducible
diagrams with genuine massive two-loop integrals \cite{Tr2} are
depicted in Figures \ref{fig3} and \ref {fig4}, respectively. Our
calculations on these diagrams are summarized in Appendix A.

\begin{figure}[h]
\begin{center}
\includegraphics{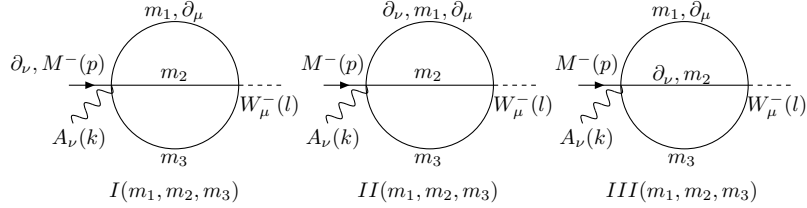}
\end{center}
\par
\vspace{2.5cm} \caption{Two-point irreducible diagrams}
\label{fig3}
\end{figure}

\begin{figure}[h]
\begin{center}
\includegraphics{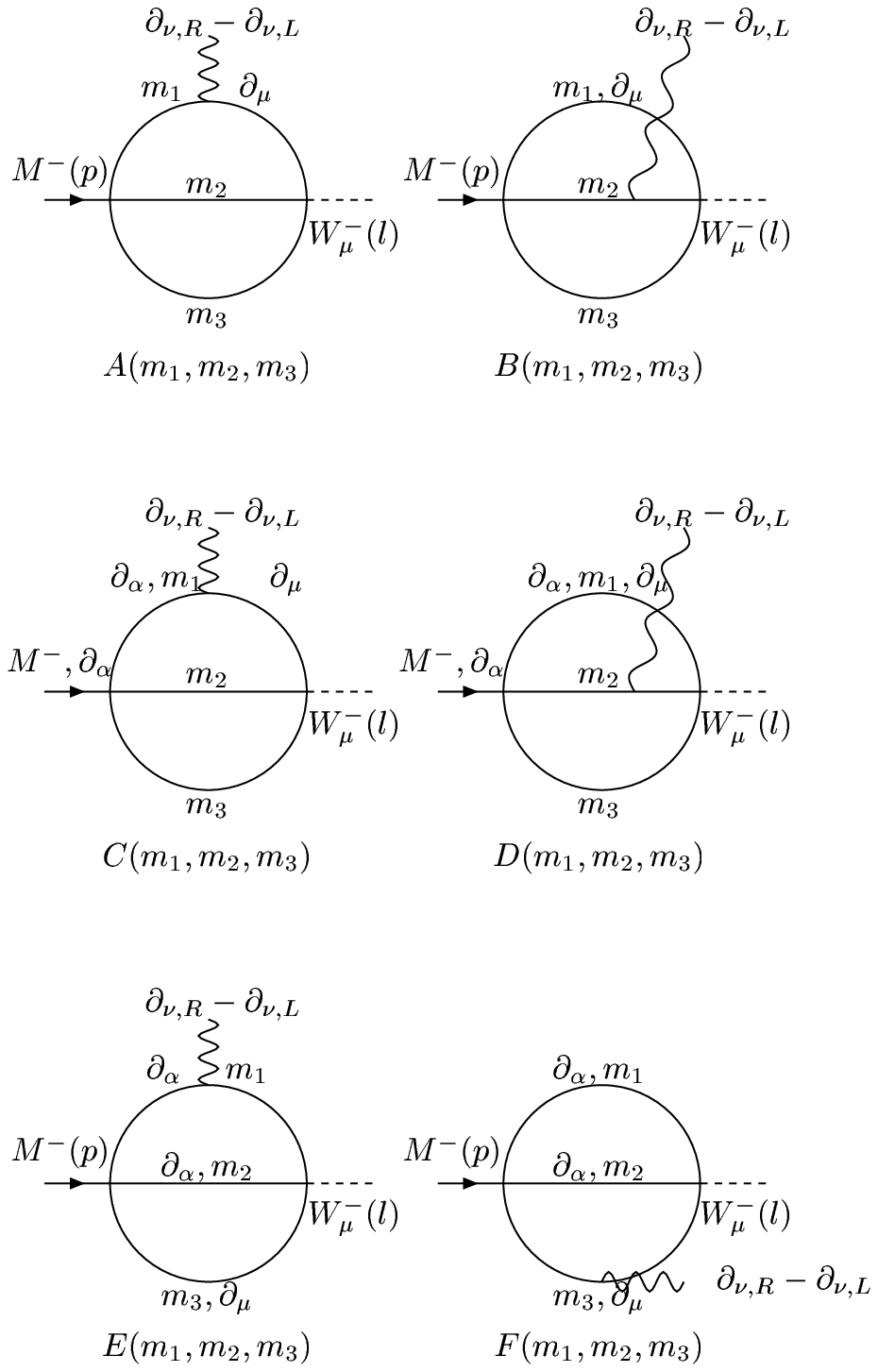}
\end{center}
\par
\vspace{8.5cm} 
\par
\begin{center}
\includegraphics{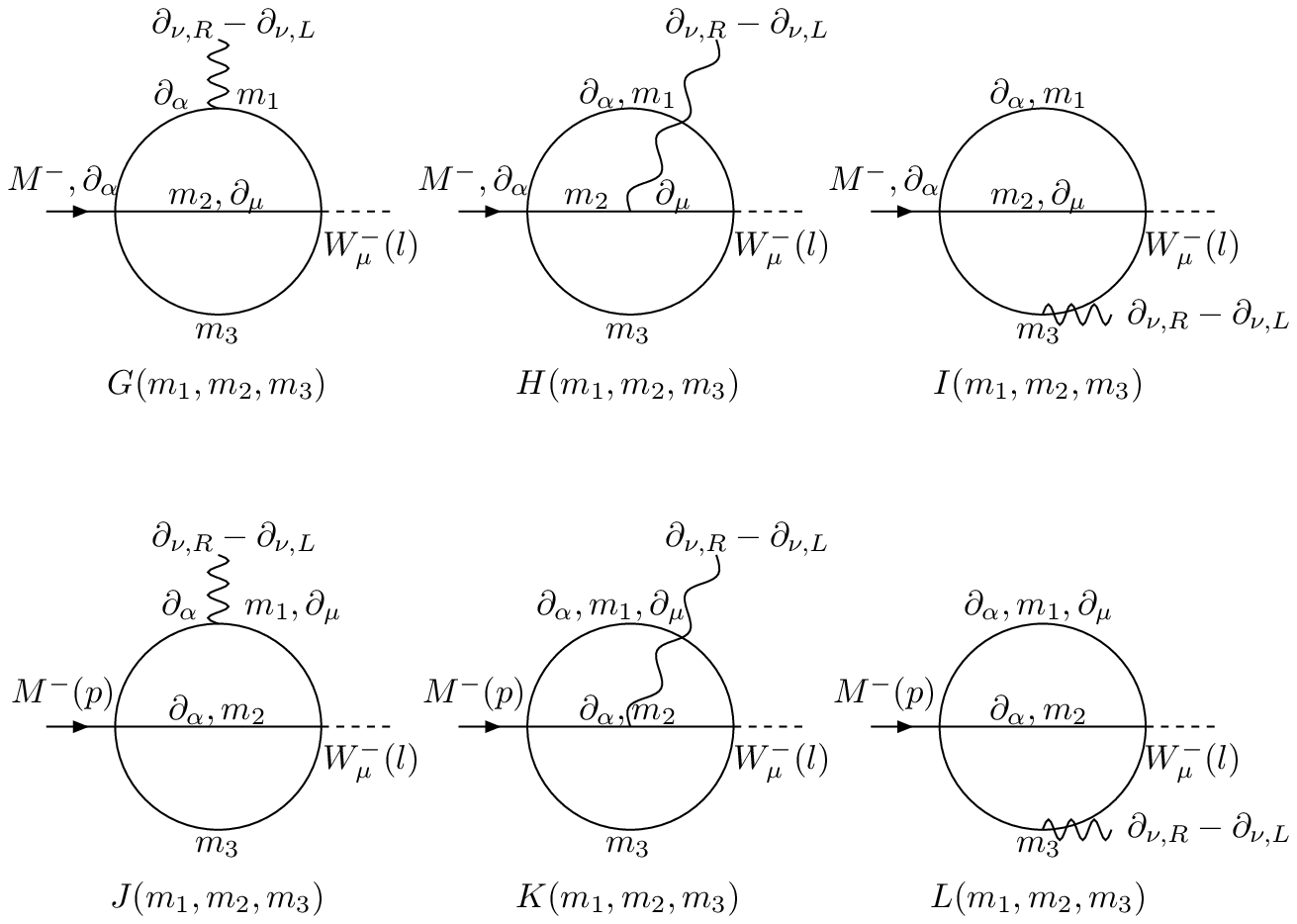}
\end{center}
\par
\vspace{5.cm} \caption{ Three-point irreducible diagrams}
\label{fig4}
\end{figure}
We point out that the two-point irreducible diagrams in Figure
\ref{fig3} vanished in Ref. \cite{BT} for $\pi_{\ell 2\gamma}$ in
the linear sigma model parametrization in which there is neither a
photon nor a four-pion vertex \cite{Bij}. As shown in Sec. 5,
these diagrams play a very important role in the cancellation of
divergent parts with $1/\epsilon^2$ in the final results.

We note that the two-loop amplitudes shown in Appendix A have been
classified by the functions $\{I,II...,A,B...L\}$ related to each
diagram. For clarity, we always refer to all inward particles for
vertices. The Feynman diagrams which contain tensor structures at
the numerator can be reduced to the calculation of scalar
integrals $P_{\alpha 1,\alpha 2,\alpha 3}^{ab}$ by
introducing the transverse components of the loop momenta $s$ and $q$ as $%
s_{\perp }^{\mu }=s^{\mu }-(s\cdot l/l^2)l^{\mu}$ and $q_{\perp
}^{\mu }=q^{\mu }-(q\cdot l/l^2)l^{\mu }$, respectively, for the
two-point diagrams \cite{Tr2}. For the three-point diagrams, we
first combine the denominators by using the Feynman formula
$1/ab=\int_{0}^{1}dx/[ax+b(1-x)]^2$ and then replace $l$ by
$l^{\prime }=l+kx$, i.e., $s_{\perp }^{\mu }=s^{\mu }-(s\cdot
l^{\prime }/l^{\prime 2})l^{\prime \mu }$
and $q_{\perp }^{\mu }=q^{\mu }-(q\cdot l^{\prime }/l^{\prime 2})%
l^{\prime \mu }$. The corresponding relations among
$\{I,II...,A,B...L\}$ and $P_{\alpha 1,\alpha 2,\alpha 3}^{ab}$
are displayed in Appendices B and C , where the scalar integrals
$P_{\alpha 1,\alpha 2,\alpha 3}^{ab}$, defined in the Euclidian
space, are given by

\begin{equation}
P_{\alpha 1,\alpha 2,\alpha 3}^{ab}(m_{1},m_{2},m_{3};l^{2})=\int
d^{n}sd^{n}q\frac{(s\cdot
l)^{a}(q.l)^{b}}{(s^{2}+m_{1}^{2})^{\alpha
1}(q^{2}+m_{2}^{2})^{\alpha 2}((s+q)^{2}+m_{3}^{2})^{\alpha 1}}\,.
\end{equation}
The detailed contributions from all diagrams for $F_{A}$ in $\pi
_{\ell 2\gamma }$ and $K_{\ell 2\gamma }$ are summarized in the
next section.

\section{Analytical Results}

\subsection{Renormalization scheme}

In our calculations, we use the following dimensional
regularization and  renormalization scheme \cite{GL1,GL2,BCE2,PS}.
Each diagram of $O(p^{2n})$ is multiplied by a factor $(c\mu
)^{(n-1)(D-4)}$, where $D=4-2\epsilon $ is the dimension of
space-time and $c$ is given by

\begin{equation}
\ln c=-\frac{1}{2}\left[1-\gamma +\ln (4\pi )\right]-\frac{\epsilon }{2}\left( \frac{%
\pi ^{2}}{12}+\frac{1}{2}\right) +O(\epsilon ^{2})\,. \label{Lnc}
\end{equation}
Using the renormalization factor of $(c\mu )^{(n-1)(D-4)}$, the
low energy constants $L _{i}$ of ${\cal L}_{n}^{(4)}$ in Eq.
(\ref{L4}) are defined by

\begin{equation}
L _{i}=(c\mu )^{(D-4)}L _{i}(\mu ,D)\,.
\end{equation}
Similarly, for the ${\cal L}_{n}^{(6)}$ parameters $y_{i}$ in Eq.
(\ref{L6}),
\begin{equation}
y_{i}=(c\mu )^{2(D-4)}y_{i}(\mu ,D)\,.
\end{equation}
We note that $L _{i}(\mu ,D)$ have the same $\mu $-dependences as
the one-loop integrals, whereas $y_{i}(\mu ,D)$ behave like the
two-loop ones. Their values at the two different scales of $\mu
_{1}$ and $\mu _{2}$ are related by $L _{i}(\mu _{1},D)=(\mu
_2/\mu _1)^{(D-4)}L _{i}(\mu _{2},D)$ and $y_{i}(\mu _{1},D)=(\mu
_2/\mu _1)^{2(D-4)}y_{i}(\mu _{2},D)$, respectively.

\subsection{Analytical forms of $F_A$}

We now try to obtain the analytical forms of $F_{A}$ from each
diagram to $O(p^{6})$ at the scale of $m_{\rho }$. From Sec. 5.1,
for the unrenormalized coefficients in the chiral lagrangian
${\cal L}_{n}^{(4)}$, we use \cite{BCE2,PS}
\begin{eqnarray}
L _{i} &=&(c\mu )^{(D-4)}\left[-\frac{\gamma _{i}}{32\pi ^{2}\epsilon }%
+L _{i}^{r}(\mu )\right]  \nonumber \\ %
&=&\mu^{(D-4)}\left\{-\frac{\gamma _{i}}{32\pi
^{2}}\left[\frac{1}{\epsilon }-\gamma _{s}-\ln \left(\frac{m_{\rho
}^{2}}{\mu ^{2}}\right)\right]+L _{i}^{r}(m_{\rho })+\epsilon L
_{i}^{(\epsilon )}(m_{\rho },\mu )\right\}\,,%
\label{alphai}
\end{eqnarray}
which is a Laurent series expanded around $\epsilon =0$. Of the
values in Eq. (\ref{alphai}), $\gamma _{i}$ are shown in Table
\ref{5.1} \cite{BCE2}, $\gamma _{s}=-1-\ln (4\pi )+\gamma $, $L
_{i}^{r}$ correspond to measurable low energy constants  and $L
_{i}^{(\epsilon )}$ are given by $L _{i}^{(\epsilon )}(m_{\rho
},\mu )=-L _{i}^{r}(m_{\rho })(\gamma _{s}+\ln (m_{\rho}^2/\mu
^2))+\gamma _{i}\cdot f(m_{\rho },\mu )$.
\begin{table}[h]
\caption{Coefficients of $\gamma _{i}$ in the Minkowski space
\cite{BCE2}. } \label{5.1}
\begin{center}
\begin{tabular}{|c|c|c|c|c|c|c|c|c|c|c|}
\hline
$i$ & $1$ & $2$ & $3$ & $4$ & $5$ & $6$ & $7$ & $8$ & $9$ & $10$ \\ \hline
$\gamma _{i}$ & $\frac{3}{32}$ & $\frac{3}{16}$ & $0$ & $\frac{1}{8}$ & $%
\frac{3}{8}$ & $\frac{11}{144}$ & $0$ & $\frac{5}{48}$ & $\frac{1}{4}$ & $%
\frac{-1}{4}$ \\ \hline
\end{tabular}
\end{center}
\end{table}
Similarly, for the chiral lagrangian ${\cal L}_{n}^{(6)}$, we have

\begin{eqnarray}
y_{i} =\frac{(c\mu )^{2(D-4)}}{F^{2}}\left\{ y_{i}^{r}(\mu
)+\left[
\Gamma_{i}^{(1)}+\Gamma _{i}^{(L)}(\mu )\right] \frac{1}{32\pi ^{2}\epsilon }%
-\Gamma _{i}^{(2)}\frac{1}{32^{2}\pi ^{4}\epsilon ^{2}}\right\} \, %
~~~~~~~~~~~~~~~~~~~~~~
 \nonumber
\\
=\frac{\mu ^{2(D-4)}}{F^{2}}\left\{ y_{i}^{r}(m_{\rho
})+\frac{\left(
\Gamma _{i}^{(1)}+\Gamma _{i}^{(L)}(m_{\rho })\right) }{32\pi ^{2}}\left[\frac{1}{%
\epsilon }-2\gamma _{s}-2\ln \left(\frac{m_{\rho }^{2}}{\mu
^{2}}\right)\right]-\Gamma
_{i}^{(2)}\frac{g(m_{\rho },\mu ,\epsilon )}{32^{2}\pi ^{4}\epsilon ^{2}}%
\right\}\,;
\end{eqnarray}
 the relevant constants of $\Gamma _{i}^{(1)}$, $\Gamma
_{i}^{(2)}$ and
$\Gamma _{i}^{(L)}$ are shown in Table \ref{5.2} \cite{BCE2}. We note that $%
f(m_{\rho },\mu )$ and $g(m_{\rho },\mu ,\epsilon )$, which
receive contributions from the $(\epsilon/2)$ term in Eq.
(\ref{Lnc}), will not contribute to $F_{A}$.

\begin{table}[h]
\caption{Coefficients of $\Gamma _{i}^{(2)}$ and  $\Gamma
_{i}^{(1,L)}$ with the double-pole and  single-pole divergences,
respectively, in the Minkowski space \cite{BCE2}. } \label{5.2}
\begin{center}
$
\begin{tabular}{|c|c|c|c|}
\hline
$y_{i}$ & $\Gamma _{i}^{(2)}$ & $16\pi ^{2}\Gamma _{i}^{(1)}$ & $\Gamma
_{i}^{(L)}$ \\ \hline
$17$ & $\frac{19}{64}$ & $-\frac{13}{768}$ & $\frac{2}{3}L _{1}^{r}+%
\frac{4}{3}L _{2}^{r}+\frac{8}{9}L _{3}^{r}+\frac{3}{4}\L
_{5}^{r}$ \\ \hline
$18$ & $\frac{67}{192}$ & $-\frac{1}{2304}$ & $\frac{8}{3}L _{1}^{r}+%
\frac{8}{9}L _{2}^{r}+\frac{23}{27}L _{3}^{r}+\frac{3}{2}\L
_{4}^{r}+\frac{1}{4}L _{5}^{r}$ \\ \hline $81$ & $0$ & $0$ &
$-\frac{3}{4}L _{9}^{r}-\frac{3}{4}L _{10}^{r}$
\\ \hline
$82$ & $0$ & $0$ & $-\frac{1}{4}L _{9}^{r}-\frac{1}{4}\L
_{10}^{r}$
\\ \hline
$83$ & $-\frac{1}{4}$ & $\frac{5}{384}$ & $\frac{1}{3}L _{1}^{r}-\frac{%
11}{6}L _{2}^{r}-\frac{5}{36}L _{3}^{r}-\frac{3}{4}L _{5}^{r}+%
\frac{3}{8}L _{9}^{r}$ \\ \hline
$84$ & $-\frac{4}{3}$ & $-\frac{5}{144}$ & $-\frac{32}{3}L _{1}^{r}-%
\frac{32}{9}L _{2}^{r}-\frac{157}{54}L _{3}^{r}-6L _{4}^{r}-\L
_{5}^{r}+\frac{1}{2}L _{9}^{r}$ \\ \hline
$85$ & $-\frac{1}{2}$ & $\frac{5}{192}$ & $\frac{2}{3}L _{1}^{r}-\frac{%
11}{3}L _{2}^{r}-\frac{16}{9}L _{3}^{r}-\frac{3}{2}L _{5}^{r}+%
\frac{3}{4}L _{9}^{r}$ \\ \hline
$100$ & $-\frac{5}{24}$ & $\frac{49}{1152}$ & $\frac{1}{3}L _{1}^{r}-%
\frac{1}{6}L _{2}^{r}+\frac{1}{4}L _{3}^{r}-\frac{1}{3}\L
_{4}^{r}-\frac{1}{4}L _{5}^{r}-\frac{9}{8}L _{9}^{r}$ \\ \hline
$102$ & $-\frac{25}{64}$ & $-\frac{17}{768}$ & $-\frac{2}{3}L _{1}^{r}-%
\frac{4}{3}L _{2}^{r}-\frac{8}{9}L _{3}^{r}-\frac{3}{4}\L
_{5}^{r}+\frac{3}{4}L _{10}^{r}$
\\ \hline $103$ & $-\frac{73}{192}$ & $-\frac{101}{2304}$ &
$-\frac{8}{3}L _{1}^{r}-\frac{8}{9}L _{2}^{r}-\frac{23}{27}L _{3}^{r}-\frac{3}{2}%
L _{4}^{r}-\frac{1}{4}L _{5}^{r}+\frac{1}{4}L _{10}^{r}$ \\ \hline
$104$ & $\frac{7}{96}$ & $\frac{5}{144}$ & $-\frac{1}{6}L _{1}^{r}+%
\frac{1}{12}L _{2}^{r}-\frac{1}{8}L _{3}^{r}+\frac{1}{6}\L
_{4}^{r}+\frac{1}{8}L _{5}^{r}+\frac{5}{16}L _{9}^{r}$
\\ \hline
$109$ & $-\frac{1}{16}$ & $\frac{3}{128}$ & $-\frac{1}{2}L
_{9}^{r}$ \\ \hline
$110$ & $\frac{1}{6}$ & $-\frac{31}{576}$ & $-\frac{2}{3}L _{1}^{r}+%
\frac{1}{3}L _{2}^{r}-\frac{1}{2}L _{3}^{r}+\frac{2}{3}\L
_{4}^{r}+\frac{1}{2}L _{5}^{r}+\frac{1}{4}L _{9}^{r}$ \\ \hline
\end{tabular}
$%
\end{center}
\end{table}

By writing the unrenormalized contributions to $F_{A}$ of
$O(p^{4})$
and $O(p^{6})$ diagrams as $F_{A,tree}(p^{4})$, $F_{A,tree}(p^{6})$, $%
F_{A,1-loop}(p^{6})$ and $F_{A,2-loop}(p^{6})$, respectively,
using the wave function renormalizations, we find \be
F_{A}&=&\sqrt{Z_{P}}\left[ F_{A,tree}(p^{4})(1+\delta
F_{P})+F_{A,tree}(p^{6})+F_{A,1-loop}(p^{6})+F_{A,2-loop}(p^{6})\right]
\,,%
\ee%
 which leads to
\be F_{A}&=&F_{A,tree}(p^{4})(1+\delta F_{P}+\frac{1}{2}\delta
Z_{P})+F_{A,tree}(p^{6})+F_{A,1-loop}(p^{6})+F_{A,2-loop}(p^{6})\,,
\label{UFA}
\ee%
 where $P=\pi $ or $K$. In Eq. (\ref{UFA}), for $P=\pi $, we
have
\begin{eqnarray}
F_{A,tree}(1+\delta F_{\pi }+\frac{1}{2}\delta Z_{\pi }) =\frac{4\sqrt{2}}{%
F_{\pi }}\left( L _{9}^{r}+L _{10}^{r}\right) \left[ 1-\frac{1}{%
F^{2}}\left( \frac{I(m_{K}^{2})}{3}+\frac{2I(m_{\pi
}^{2})}{3}\right) \right] ~~~~~~~~~~~~~~~  \nonumber \\
=\frac{4\sqrt{2}}{F_{\pi }}\left( L _{9}^{r}+L _{10}^{r}\right)
\left\{ 1+\frac{1}{16\pi ^{2}F_{\pi }^{2}}\left[ \frac{1}{\varepsilon }%
-2\gamma _{s}-2\ln (\frac{m_{\rho }^{2}}{\mu ^{2}})\right] \left( \frac{%
m_{K}^{2}}{3}+\frac{2m_{\pi }^{2}}{3}\right) \right.
\nonumber \\ %
\left. -\frac{1}{16\pi ^{2}F_{\pi }^{2}}\left[ \frac{m_{K}^{2}}{3}\ln \left(%
\frac{m_{K}^{2}}{m_{\rho }^{2}}\right)+\frac{2m_{\pi }^{2}}{3}\ln
\left(\frac{m_{\pi }^{2}}{m_{\rho }^{2}}\right)\right] \right\}
\,,~~~~~~~ \label{FAPT}
\end{eqnarray}
\begin{eqnarray}
F_{A,1-loop} &=&\frac{1}{6\sqrt{2}F_{\pi }^{3}\pi ^{2}}\left[ \frac{1}{%
\epsilon }-2\gamma _{s}-2\ln \left(\frac{m_{\rho }^{2}}{\mu
^{2}}\right)\right] \times %
\nonumber \\ &&\left[ 6(2L _{1}^{r}-L _{2}^{r})m_{\pi
}^{2}+(m_{K}^{2}+2m_{\pi }^{2})(3L _{3}^{r}+5L _{9}^{r}+5L
_{10}^{r})\right] \nonumber
\\
&&+\frac{1}{6\sqrt{2}F_{\pi }^{3}\pi ^{2}}\left\{ (-12L _{1}^{r}
+6L _{2}^{r})m_{\pi }^{2}\ln \left(\frac{m_{\pi }^{2}}{m_{\rho }^{2}}%
\right)-(3L _{3}^{r}+5L _{9}^{r}+5L_{10}^{r})\times \right. %
\nonumber \\ &&\left. \left[ m_{K}^{2}\ln
\left(\frac{m_{K}^{2}}{m_{\rho }^{2}}\right)+2m_{\pi }^{2}\ln
\left(\frac{m_{\pi }^{2}}{m_{\rho }^{2}}\right)\right] \right\}
\,, \label{FAP1}
\end{eqnarray}
\begin{eqnarray}
F_{A,2-loop} &=&\frac{1}{6F_{\pi }^{3}(2\pi )^{8}}\left\{ -\frac{\pi
^{4}(116m_{K}^{2}+184m_{\pi }^{2}+21m_{\eta }^{2})}{12\sqrt{2}}\left[ \frac{1%
}{\epsilon }-2\gamma _{s}-2\ln \left(\frac{m_{\rho }^{2}}{\mu
^{2}}\right)\right] \right.  \nonumber \\
&&-\frac{3\pi ^{4}}{2\sqrt{2}}\left[ \frac{1}{\epsilon }-2\gamma _{s}-2\ln \left(%
\frac{m_{\rho }^{2}}{\mu ^{2}}\right)\right] (p\cdot k)%
  \nonumber
\\ &&\left. -1421.4-1167.7(p\cdot k)+1228.0+1123.2(p\cdot k)\right\} \,,
\label{FAP2}
\end{eqnarray}
\begin{eqnarray}
F_{A,tree}(p^{6}) &=&-\left[ \frac{1}{\epsilon }-2\gamma _{s}-2\ln \left(\frac{%
m_{\rho }^{2}}{\mu ^{2}}\right)\right] \left\{ -\frac{1}{6F_{\pi }^{3}(2\pi )^{8}}%
\frac{3\pi ^{4}}{2\sqrt{2}}(p\cdot k)\right.  \nonumber \\
&&-\frac{1}{6F_{\pi }^{3}(2\pi )^{8}}\frac{\pi
^{4}(432m_{K}^{2}+531m_{\pi
}^{2})}{36\sqrt{2}}+\frac{1}{6\sqrt{2}F_{\pi }^{3}\pi ^{2}}\left[
3m_{K}^{2}(L _{3}^{r}+2L _{9}^{r}+2L _{10}^{r})\right. \nonumber
\\ &&\left. \left. +6m_{\pi }^{2}(2L _{1}^{r}-L _{2}^{r}+L
_{3}^{r}+2L _{9}^{r}+2L _{10}^{r})\right] \right\}  \nonumber \\
&&-\frac{4\sqrt{2}}{F_{\pi }^{3}}\left\{
4m_{K}^{2}(6y_{18}^{r}-2y_{82}^{r}+y_{84}^{r}+2y_{103}^{r})\right.
\nonumber
\\
&&+2m_{\pi
}^{2}(6y_{17}^{r}+6y_{18}^{r}-2y_{81}^{r}
-2y_{82}^{r}+2y_{83}^{r}+y_{84}^{r}+y_{85}^{r}-y_{100}^{r}+2y_{102}^{r}
\nonumber \\
&&\left. +2y_{103}^{r}-2y_{104}^{r}+y_{109}^{r})+pk(2y_{100}^{r}
-4y_{109}^{r}+y_{110}^{r})\right\} \,. \label{FAT6}
\end{eqnarray}
For $P=K$, we get
\begin{eqnarray}
F_{A,tree}(1+\delta F_{K}+\frac{1}{2}\delta Z_{K}) =\frac{4\sqrt{2}}{F_{K}}%
(L _{9}^{r}+L _{10}^{r})\left[ 1-\frac{1}{F^{2}}\left( \frac{%
I(m_{\eta }^{2})}{4}+\frac{I(m_{K}^{2})}{2}+\frac{I(m_{\pi }^{2})}{4}\right) %
\right] ~~  \nonumber \\
=\frac{4\sqrt{2}}{F_{K}}(L _{9}^{r}+L _{10}^{r})\left\{ 1+\frac{1%
}{16\pi ^{2}F_{K}^{2}}\left[ \frac{1}{\varepsilon }-2\gamma _{s}-2\ln\left(\frac{%
m_{\rho }^{2}}{\mu ^{2}}\right)\right] \left( \frac{m_{\eta }^{2}}{4}+\frac{%
m_{K}^{2}}{2}+\frac{m_{\pi }^{2}}{4}\right) \right.  \nonumber \\
\left. -\frac{1}{16\pi ^{2}F_{K}^{2}}\left[ \frac{m_{\eta }^{2}}{4}\ln \left(%
\frac{m_{\eta }^{2}}{m_{\rho }^{2}}\right)+\frac{m_{K}^{2}}{2}\ln \left(\frac{m_{K}^{2}%
}{m_{\rho }^{2}}\right)+\frac{m_{\pi }^{2}}{4}\ln
\left(\frac{m_{\pi }^{2}}{m_{\rho }^{2}}\right)\right] \right\}
\,,~ \label{FAKT}
\end{eqnarray}
\begin{eqnarray}
F_{A,1-loop} &=&\frac{1}{4\sqrt{2}F_{K}^{3}\pi ^{2}}\left[
\frac{1}{\epsilon }-2\gamma _{s}-2\ln \left(\frac{m_{\rho
}^{2}}{\mu ^{2}}\right)\right] \left\{ m_{K}^{2}(8L _{1}^{r}-4L
_{2}^{r}+4L _{3}^{r}+6L _{9}^{r}+6L _{10}^{r})\right. \nonumber
\\ &&\left. +m_{\pi }^{2}(2L _{3}^{r}+3L _{9}^{r}+3L _{10}^{r})
+m_{\eta }^{2}(L _{9}^{r}+L _{10}^{r})\right\}  \nonumber
\\
&&+\frac{1}{4\sqrt{2}F_{K}^{3}\pi ^{2}}\left\{ -(2L
_{3}^{r}+3L _{9}^{r}+3L _{10}^{r})m_{\pi }^{2}\ln \left(\frac{m_{\pi }^{2}}{m_{\rho }^{2}%
}\right)-(L _{9}^{r}+L _{10}^{r})m_{\eta }^{2}\ln \left(\frac{m_{\eta }^{2}}{%
m_{\rho }^{2}}\right)\right.  \nonumber \\ &&\left. -(8L
_{1}^{r}-4L _{2}^{r}+4L _{3}^{r}+6L _{9}^{r}
+6L _{10}^{r})m_{K}^{2}\ln \left(\frac{m_{K}^{2}}{m_{\rho }^{2}}%
\right)\right\} \,,  \label{FAK1}
\end{eqnarray}
\begin{eqnarray}
F_{A,2-loop} &=&\frac{1}{6F_{K}^{3}(2\pi )^{8}}\left\{ -\frac{\pi
^{4}(99m_{K}^{2}+20m_{\pi }^{2}-12m_{\eta }^{2})}{4\sqrt{2}}\left[ \frac{1}{%
\epsilon }-2\gamma _{s}-2\ln \left(\frac{m_{\rho }^{2}}{\mu
^{2}}\right)\right] \right. \nonumber \\
&&-\frac{3\pi ^{4}}{2\sqrt{2}}\left[ \frac{1}{\epsilon }-2\gamma _{s}-2\ln\left(%
\frac{m_{\rho }^{2}}{\mu ^{2}}\right)\right] (p\cdot k)  \nonumber
\\ &&\left. -1465.2-923.0(p\cdot k)+1266.9+1208.3(p\cdot k)\right\} \,,
\label{FAK2}
\end{eqnarray}
\begin{eqnarray}
F_{A,tree}(p^{6}) &=&-\left[ \frac{1}{\epsilon }-2\gamma _{s}-2\ln \left(\frac{%
m_{\rho }^{2}}{\mu ^{2}}\right)\right] \left\{ -\frac{1}{6F_{K}^{3}(2\pi )^{8}}%
\frac{3\pi ^{4}}{2\sqrt{2}}(p\cdot k)\right.  \nonumber \\
&&-\frac{1}{6F_{K}^{3}(2\pi )^{8}}\frac{\pi
^{4}(747m_{K}^{2}+216m_{\pi
}^{2})}{36\sqrt{2}}+\frac{1}{4\sqrt{2}F_{K}^{3}\pi ^{2}}\left[
m_{\pi }^{2}(2L _{3}^{r}+3L _{9}^{r}+3L _{10}^{r})\right.
\nonumber
\\
&&\left. \left. +m_{K}^{2}(8L _{1}^{r}-4L _{2}^{r}+4L _{3}^{r}+9L
_{9}^{r}+9L _{10}^{r})\right] \right\}  \nonumber \\
&&-\frac{4\sqrt{2}}{3F_{K}^{3}}\left\{ 2m_{\pi
}^{2}(18y_{18}^{r}-2y_{81}^{r}-6y_{82}^{r}
+2y_{83}^{r}+3y_{84}^{r}-y_{85}^{r}+6y_{103}^{r})\right. \nonumber
\\ &&+2m_{K}^{2}(18y_{17}^{r}+36y_{18}^{r}-4y_{81}^{r}
-12y_{82}^{r}+4y_{83}^{r}+6y_{84}^{r}+4y_{85}^{r}-3y_{100}^{r}
\nonumber \\ &&\left.
+6y_{102}^{r}+12y_{103}^{r}-6y_{104}^{r}+3y_{109}^{r})
+3pk(2y_{100}^{r}-4y_{109}^{r}+y_{110}^{r})\right\} \,.
\label{FAK6}
\end{eqnarray}

We note that, in the above expressions, one special case must be treated
separately. For the finite part of $F_{A,2-loop}$, the functions of $g(y)$ and $%
f_{i}(y)$ in Appendix B seem to introduce additional singularities
in the integrals, such as at $y=0$ and $1$. However, this problem
can be resolved by noticing that at $y=0$ (the situation is the
same at y=1), the function, e.g. $g(y)$, behaves like $\ln
(x^{2})$, which is integrable. Consequently, the main question of
evaluating  the finite part of $F_{A,2-loop}$ is how to implement
the formula in a computer program by correct and numerically
stable forms. As seen from the last two terms for $F_{A,2-loop}$
in Eqs. (\ref{FAP2}) and (\ref{FAK2}), we have found the reliable
and stable numerical results due to the contributions of functions
$h_{i}$, i.e., the integrals of $g$ and $f_{i}$, by fitting the
numerical data with recursive analytic methods.

We remark that in Eqs. (\ref{FAPT})-(\ref{FAK6}) we have
explicitly shown the single poles, subtracted via
$F_{A,tree}(p^{6})$. We emphasize that in our results there are no
divergent parts with $1/\epsilon ^{2}$ and all the terms related
to $1/\epsilon $ are cancelled explicitly by the renormalization
of the coupling constants in ${\cal L}_{n}^{(6)}$ as well as the
Gell-Mann-Okubo relation in Eq. (\ref{GO}). It is clear that the
disappearance of $1/\epsilon ^{2}$ terms relies on  the two-point
irreducible diagrams in Figure \ref{fig3}.
Moreover, our results are scale independent since the scale terms
with $\ln \mu ^{2}$ can be grouped into the ones with $1/\epsilon
$, i.e., they are always associated with  $1/\epsilon $ terms.
These also serve as checks of our calculations. Now we can extract
the axial-vector form factor by placing the related physical
quantities into the Eq. (\ref{UFA}).
Explicitly, we may write Eq.
(\ref{UFA}) into more transparent forms

\begin{eqnarray}
F_{A,\pi } &=&\left\{ 66.86(L_{9}^{r}+L_{10}^{r})\right\}
\nonumber
\\
&&+\{(2.41-122.96(pk))\frac{y_{100}^{r}}{F^{2}}-4.82\frac{y_{102}^{r}}{F^{2}}%
-125.35\frac{y_{103}^{r}}{F^{2}}+4.82\frac{y_{104}^{r}}{F^{2}}\nonumber
\\
&&+(-2.41+245.95(pk))\frac{y_{109}^{r}}{F^{2}}-61.49(pk)\frac{y_{110}^{r}}{%
F^{2}}-14.46\frac{y_{17}^{r}}{F^{2}}-376.05\frac{y_{18}^{r}}{F^{2}}\nonumber
\\
&&+4.82\frac{y_{81}^{r}}{F^{2}}+125.35\frac{y_{82}^{r}}{F^{2}}-4.82\frac{%
y_{83}^{r}}{F^{2}}-62.67\frac{y_{84}^{r}}{F^{2}}-2.41\frac{y_{85}^{r}}{F^{2}}%
\}  \nonumber \\ &&+\left\{
12.30L_{1}^{r}-6.15L_{2}^{r}+16.11L_{3}^{r}+26.85L_{9}^{r}+26.85L_{10}^{r}%
\right\}   \nonumber \\ &&+\left\{ -1.70\cdot 10^{-2}-3.92\cdot
10^{-3}(pk)\right\} \label{n55}
\end{eqnarray}
and
\begin{eqnarray}
F_{A,K} &=&\left\{ 54.99(L_{9}^{r}+L_{10}^{r})\right\}   \nonumber
\\
&&+\{(24.75-101.01(pk))\frac{y_{100}^{r}}{F^{2}}-49.50\frac{y_{102}^{r}}{%
F^{2}}-102.97\frac{y_{103}^{r}}{F^{2}}+49.50\frac{y_{104}^{r}}{F^{2}}
\nonumber
\\
&&+(-24.75+202.03(pk))\frac{y_{109}^{r}}{F^{2}}-50.50(pk)\frac{y_{110}^{r}}{%
F^{2}}-148.50\frac{y_{17}^{r}}{F^{2}}-308.90\frac{y_{18}^{r}}{F^{2}}
\nonumber
\\
&&+34.32\frac{y_{81}^{r}}{F^{2}}+102.97\frac{y_{82}^{r}}{F^{2}}-34.32\frac{%
y_{83}^{r}}{F^{2}}-51.48\frac{y_{84}^{r}}{F^{2}}-32.34\frac{y_{85}^{r}}{F^{2}%
}\}  \nonumber \\ &&+\left\{
22.08L_{1}^{r}-11.04L_{2}^{r}+12.75L_{3}^{r}+21.71L_{9}^{r}+21.71L_{10}^{r}%
\right\}   \nonumber \\ &&+\left\{ -0.97\cdot 10^{-2}+13.93\cdot
10^{-3}(pk)\right\}\,, \label{n56}
\end{eqnarray}
where the four $\{\cdots\}$ terms in Eqs. (\ref{n55}) and
(\ref{n56}) correspond to those in Eq. (\ref{UFA}), respectively.

\section{Numerical values and conclusions}

As shown in section 5, the divergent terms for $F_A$ in
loop-diagrams are cancelled by the corresponding counterterms in
the Lagrangian at $O(p^{6})$. The infinite parts cancel each other
and thus they can be simply substituted by the remaining finite
part of the counterterms, $y_{i}^{r}$. We now study the finite
parts which contain the actual physical information. We will
present the results in numerical forms, with the scale at $m_{\rho
}=0.77\ GeV$. In Table  \ref{6.1}, we show the standard values for
the couplings $L _{i}^{r} $ in ${\cal L}_{n}^{(4)}$
\cite{Handbook} and those in the two-loop calculation of ChPT; we
chose the Main Fit in Ref. \cite{Refs} as an illustration. Other
two-loop studies in ChPT can be found in Refs. \cite{Others,Sc}.
We note that in the table the central value of
$\alpha_{10}^r=-5.5$ is kept, and our numerical results for $F_A$
are sensitive to this value.
 Several $O(p^{6})$ low-energy constants of the normal chiral
Lagrangian have been evaluated from the RS.
In Table \ref{6.3} we illustrate the values of $y_i$ in the lowest
meson dominance (LMD) approximation \cite{KN} and the
 resonance Lagrangian (RL)
 \cite{ABBC,KN,Pr}.
\begin{table}[h]
\caption{Values of $L _{i}^{r}$ in ${\cal L}_{n}^{(4)}$. }
\label{6.1}
\begin{center}
\begin{tabular}{|c|c|}
\hline $10^{3}L _{i}^{r}$ & $1$ \ \ \ \ \ \ \ \ \ \ \ \ \ \ $2$ \
\ \ \ \ \ \ \ \ \ \ \ \ \ $3$ \ \ \ \ \ \ \ \ \ \ \ \ \ \ $9$ \ \
\ \ \ \ \ \ \ \ \ \ \ \ $10$ \\ \hline $O(p^{4})$ \cite{Handbook}
& $0.4\pm 0.3$ \ \ \ \ $1.35\pm 0.30$ \ \ \ \ $-3.5\pm 1.1 $ \ \ \
\ $6.9\pm 0.7$ \ \ \ $-5.5\pm 0.7$
\\ \hline Main Fit \cite{Refs} & $0.53\pm 0.25$ \ \
$0.71\pm 0.27$ \ \  $-2.72\pm 1.12 $ \ \ $6.9\pm 0.7$ \ \ $-5.5\pm
0.7$
\\ \hline
\end{tabular}
\end{center}
\end{table}

\begin{table}[h]
\caption{Values of $y_{i}$ in ${\cal L}_{n}^{(6)}$.} \label{6.3}
\begin{center}
\begin{tabular}{|c|c|c|c|c|}
\hline $y_{i}(in$ $units$ $of$ $10^{-4}/F^{2})$ & $y_{100}$ &
$y_{104}$ & $y_{109}$ & $y_{110}$%
\\ \hline $LMD$ & $1.09$ & $-0.36$ &$0.40$ & $-0.52$%
\\ \hline $RL$ $I$ & $1.09$ &$-0.29$ & $0.47$ & $-0.16$
\\ \hline $RL$ $II$ & $1.49$ & $-0.39$ &$0.65$ & $-0.14$%
 \\ \hline
\end{tabular}
\end{center}
\end{table}

To  study the vector form factors, we need to consider the
anomalous chiral Lagrangian. The set of anomalous coefficients is
treated by phenomenological fitting in ChPT as well as by the two
main alternative models of vector meson dominance (VMD) method and
constituent chiral quark model (CQM) \cite{St}. The relevant terms
for our purposes are shown in Table \ref{6.2}.
\begin{table}[h]
\caption{Values of  $C_{i}^{Wr}$ in ${\cal L}_{a}^{(6)}$ in
various models \cite{St}. } \label{6.2}
\begin{center}
\begin{tabular}{|c|c|c|c|}
\hline $C_{i}^{Wr}[10^{-3}GeV^{-2}]$ & $7$ & $11$ & $22$ \\ \hline
$ChPT$ & $
\begin{array}{c}
0.013\pm 1.17 \\ 20.3\pm 18.7
\end{array}
$ & $-6.37\pm 4.54$ & $
\begin{array}{c}
6.52\pm 0.78 \\ 5.07\pm 0.71
\end{array}
$ \\ \hline
$VMD$ &  &  & $\frac{3}{64M_{\rho }^{2}\pi ^{2}}\simeq
8.01$
\\ \hline
$CQM$ & $0.51\pm 0.06$ & $-0.00143\pm 0.03$ & $3.94\pm 0.43$
\\ \hline
\end{tabular}
\end{center}
\end{table}
 Other physical inputs are
$m_{K}=0.495$, $m_{\pi }=0.14$, $m_{\eta }=0.55$, $F_{K}=0.112$,
$F=0.0871$ and $F_{\pi }=0.092\ GeV$. We note that some of the
actual values of $F_\pi$ and the masses in our calculations may be
different from those in the literature. In particular $F_\pi$
could differ around $1\%$ from one paper to another; however, we
expect that the changes on our $O(p^6)$ results due to the
different sets of parameters are less than $5\%$ since $O(p^6)$
contributions are at least proportional to $F_\pi^3$.

 To compare our results with
those in the literature, we use dimensionless form factors of
$f_{V,A}$, defined by \be
f_i &=& m_P F_i\,,\ \ \ (i=V,A) %
\ee %
to replace $F_{V,A}$.

\begin{figure}[h]
\vspace{1cm}
\par
\begin{center}
\includegraphics{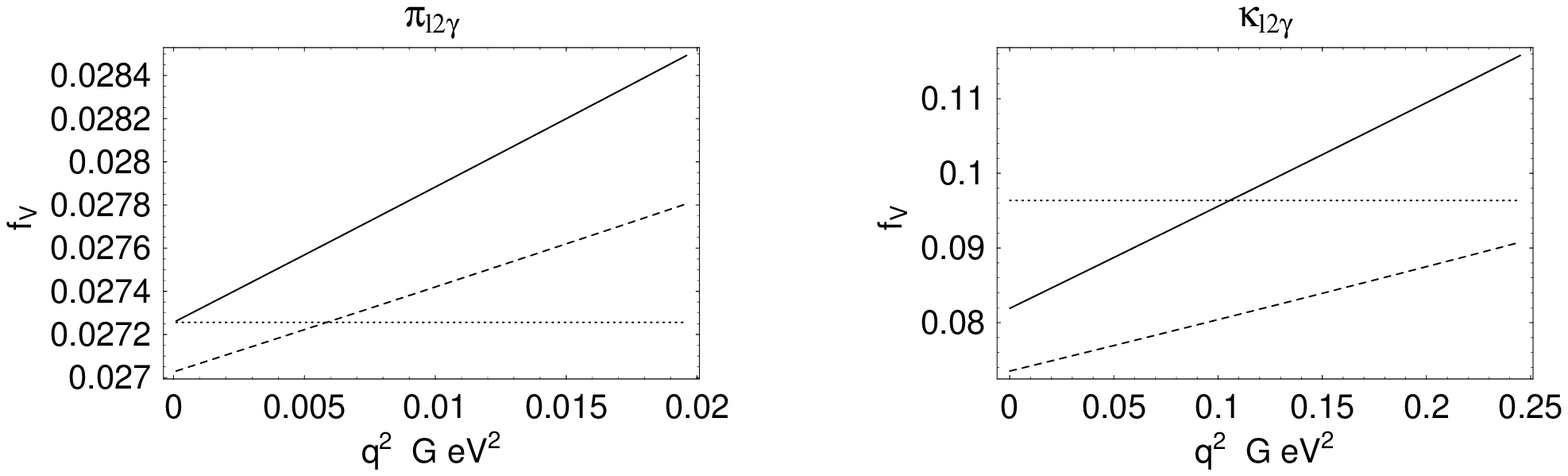}
\end{center}
\par
\vspace{3.5cm} \caption{The vector form factors $f_V$ as functions
of the momentum transfer $q^2$ for $\protect\pi_{\ell
2\protect\gamma}$ and $K_{\ell 2\protect\gamma}$ with $\ell=e$.
The dot, solid and dashed curves stand for the contributions of
$O(p^4)$, $O(p^4)+O(p^6)$ in VMD and $O(p^4)+O(p^6)$ in CQM,
respectively. } \label{fig5}
\end{figure}

\begin{figure}[h]
\vspace{3.cm}
\par
\begin{center}
\includegraphics{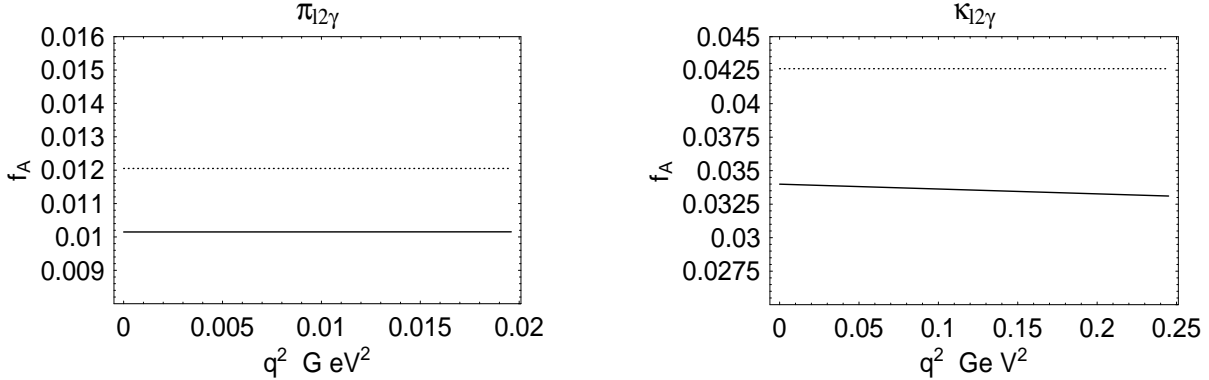}
\end{center}
\par
\vspace{2.5cm}
\caption{Same as Figure \protect\ref{fig5} but for the axial-vector form factors $%
f_A $. The dashed and solid curves represent the contributions at
$O(p^4)$ in Table 3 and the fitting $O(p^4)+O(p^6)$, respectively.
} \label{fig6}
\end{figure}

 In Figures \ref{fig5} and \ref{fig6}, we plot the dimensionless vector
and axial-vector form factors $f_{V,A}$ as functions of $q^{2}$
with the photon on mass-shell for $\pi _{e2\gamma }$ and $K
_{e2\gamma }$, respectively. Similar figures can also be drawn for
the $\mu $ modes. In Table \ref{6.4}, we show the form factors of
$f_A$ at $q^2=0$ at $O(p^4)$ and $O(p^6)$ with $SU(2)$ and $SU(3)$
symmetries as well as experimental values for $P=K$ and $\pi$.

\begin{table}[h]
\caption{  $f_{A}$ at $q^2=0$ for $P=K$ and $\pi$.} \label{6.4}
\begin{center}
\begin{tabular}{|l|l|l|l|l|}
\hline $f_{A}(q^2=0)$  & $O(p^{4})$ \cite{Handbook} &
$O(p^{6})|_{SU(2)}$  & $O(p^{6})|_{SU(3)}$ & $Experiment$
\\ \hline $P=K$ & $0.041$ & $- $ & $0.034$ & $0.035\pm 0.020$ \cite{expt,BGK} \\
\hline $P=\pi $ & $0.0102$ & $0.0117$ \cite{BT}& $0.0112$ &
$0.0116\pm 0.0016$ \cite{PDG}\\ \hline
\end{tabular}
\end{center}
\end{table}

In Figure \ref{fig5}, the dot, solid
and dashed curves stand for the contributions to $f_V$ at $O(p^{4})$, $%
O(p^{4})+O(p^{6})$ in VMD and $O(p^{4})+O(p^{6})$ in CQM,
respectively. We note that for $K_{l2\gamma}$ in Figure \ref{fig5}
$F$ has been set to $F_\pi$ in Eq. (\ref{n29}) for the curve of
$O(p^4)$ as in the literature and $F_K$ in Eq. (\ref{FVK}) for
those of $O(p^4)+O(p^6)$.
As shown in Figure \ref{fig5}, the $O(p^{6})$ contribution
obtained for the $\pi $ radiative decays is very small ($<5\%$)
for all the kinematical allowed values. However, it is interesting
to see that
 the $O(p^{6})$ correction for $K_{\ell 2\gamma }$ is much larger.

In Figure \ref{fig6}, for $f_{A}$, the dashed and solid curves
represent the contributions at  $O(p^{4})$ %
in Table 3 plus the two-loop calculations using either the
 LMD or the RL determinations, respectively.
It is easy to see that, as shown in the figures, the two-loop
contributions to $f_A$  are sizable and destructive compared with
those at the pure $O(p^4)$ for both $\pi$ and $K$ modes, but their
$q^{2}$-dependences, which are dominated by the irreducible
diagrams, are small. At $q^2=0$, the  contributions to $f_A$ from
$f_{A,tree}(p^6)$ are vanishingly small, which implies that the
final results of $f_A$ are insensitive to the known values of
$y_i^r$. However, those from the irreducible two-loop diagrams and
the one-loop diagrams with one vertex of ${\cal L}_{n}^{(4)}$ give
the dominant corrections to $f_A$ at $O(p^4)$. All together the
$O(p^6)$ corrections keep around $25\%$ for both decays. We remark
that the uncertainties in Eqs. (\ref{n55}) and (\ref{n56}) due to
the errors of $y_i^r$ \cite{KN} are less than $1\%$. Moreover, our
results of tree contributions at $O(p^6)$ are also insensitive to
the choice of the scale, as expected.

We note that, as shown in Figure \ref{fig5}, the numerical result
for $\pi \rightarrow e\nu _{e}\gamma $ at $O(p^6)$, using the
$SU(3)\otimes SU(3)$ chiral symmetry, is found to be  comparable
to the one in Ref. \cite{BT} from resonance estimates of
$O(p^{6})$ low-energy constants based on $SU(2)\otimes SU(2)$.
Furthermore, our result of the $O(p^6)$ correction for $f_A$ in
$K_{l2\gamma}$ also confirms the speculation in Ref. \cite{BT}.

In summary, we have studied the $O(p^{6})$ corrections to the
vector and axial-vector form factors in $\pi _{\ell 2\gamma }$ and
$K_{\ell 2\gamma }$ decays. These include the contributions from
loop diagrams and the ones from higher dimension terms in the
lagrangians. The former can be exactly calculated in terms of the
known parameters of the chiral lagrangians. The latter is mainly
evaluated from the resonance contributions. For the axial-vector
form factors of $f_A$, we have found that the divergent parts
cancel order by order in ChPT, while the finite numerical results
in both $K$ and $\pi$ modes contain considerable corrections from
loops diagrams; they also agree with the recent experimental
determination \cite{expt,BGK}. This demonstrates numerically the
statement about the final-state theorem mentioned in Ref. \cite
{Tr1,DHT,Tr2}.
Finally, we remark that our result of $f_A$ at $q^2$=0 for the
kaon case is consistent with that found in the light front QCD
model \cite{geng1,geng2} as well.

\section*{Acknowledgments}
We would like to thank Drs. Bijnens and Talavera for encouragement
and comments. We would also like to thank Elizabeth Walter for
reading the manuscript. This work was supported in part by the
National Science Council of the Republic of China under contract
numbers
 NSC-91-2112-M-007-043.

\newpage

\section*{Appendix A}

From Figure \ref{fig3}, we have
\begin{eqnarray}
M_{A,\pi,2-point } &=&\frac{\sqrt{2}eG_{F}l^{\mu }\varepsilon
^{\nu }}{12\times 24F^{3}}\cos \theta
[-40\sqrt{2}I(m_{K},m_{K},m_{\pi })+40\sqrt{2}I(m_{\pi
},m_{K},m_{K})  \nonumber \\
&&+24\sqrt{2}I(m_{K},m_{K},m_{\eta })-24\sqrt{2}I(m_{\eta },m_{K},m_{K}) -240%
\sqrt{2}II(m_{K},m_{\pi },m_{K})  \nonumber \\
&&
-40\sqrt{2}II(m_{\pi },m_{K},m_{K})
-320\sqrt{2}II(m_{\pi},m_{\pi },m_{\pi })  
\nonumber \\
&&-48\sqrt{2}II(m_{K},m_{K},m_{\eta })-72\sqrt{2}II(m_{\eta},m_{K},m_{K}) 
\nonumber \\ 
&&+240\sqrt{2}III(m_{K},m_{K},m_{\pi})+
40\sqrt{2}III(m_{K},m_{\pi },m_{K})
 \nonumber \\
&&+320\sqrt{2}III(m_{\pi },m_{\pi },m_{\pi })
+96\sqrt{2}III(m_{\eta},m_{K},m_{K})  
\nonumber \\
&&+72\sqrt{2}III(m_{K},m_{\eta},m_{K})
-48\sqrt{2}III(m_K,m_K,m_{\eta})],
 \\ 
\nonumber \\ 
M_{A,K,2-point} &=&\frac{\sqrt{2}eG_{F}l^{\mu }\varepsilon ^{\nu
}}{12\times 24F^{3}}\sin
\theta [-36\sqrt{2}I(m_{\eta },m_{\eta },m_{K})+36\sqrt{2}%
I(m_{K},m_{\eta },m_{\eta })  \nonumber \\
&&-36\sqrt{2}I(m_{\pi },m_{\pi},m_{K})+36\sqrt{2}I(m_{K},m_{\pi
},m_{\pi })
\nonumber \\
&&-36\sqrt{2}II(m_{K},m_{\eta },m_{\eta
})-336\sqrt{2}II(m_{K},m_{K},m_{K})
\nonumber \\
&&-36\sqrt{2}II(m_{K},m_{\pi },m_{\pi }) -168\sqrt{2}II(m_{\pi
},m_{\pi
},m_{K})  \nonumber \\
&&-36\sqrt{2}II(m_{\eta },m_{K},m_{\pi
})-96\sqrt{2}II(m_{K},m_{\pi
},m_{\eta })  \nonumber \\
&&-12\sqrt{2}II(m_{\pi },m_{K},m_{\eta })+36\sqrt{2}III(m_{\eta
},m_{K},m_{\eta })  \nonumber \\
&&+336\sqrt{2}III(m_{K},m_{K},m_{K})+168\sqrt{2}III(m_{\pi
},m_{\pi},m_{K})
\nonumber \\
&&+36\sqrt{2}III(m_{\pi },m_{K},m_{\pi })-36\sqrt{2}III(m_{\pi
},m_{\eta},m_{K})  \nonumber \\
&&+72\sqrt{2}III(m_{K},m_{\eta },m_{\pi })+48\sqrt{2}III(m_{\eta
},m_{K},m_{\pi })  \nonumber \\
&&+48\sqrt{2}III(m_{\pi },m_{K},m_{\eta })
+24\sqrt{2}III(m_{K},m_{\pi
},m_{\eta })  \nonumber \\
&&-12\sqrt{2}III(m_{\eta },m_{\pi },m_{K})].
\end{eqnarray}

\noindent
 Form Figure \ref{fig4}, we obtain
\begin{eqnarray}
M_{A,K,3-point} &=&\frac{i\sqrt{2}}{12F^{3}}eG_{F}l^{\mu
}\varepsilon ^{\nu }\sin \theta
\{6\sqrt{2}m_{K}^{2}A(m_{K},m_{K},m_{K}) \nonumber \\
&&+\sqrt{2}(m_{K}^{2}+m_{\pi }^{2})A(m_{\pi },m_{\pi },m_{K})+\frac{\sqrt{2}%
}{2}(m_{K}^{2}+m_{\pi }^{2})A(m_{K},m_{\pi },m_{\pi })  \nonumber \\
&&+\frac{\sqrt{2}}{2}(3m_{K}^{2}-m_{\pi }^{2})A(m_{K},m_{\eta },m_{\eta })+%
\frac{\sqrt{2}}{3}(m_{K}^{2}-m_{\pi }^{2})A(m_{\pi },m_{\eta
},m_{K})
\nonumber \\
&&-\frac{\sqrt{2}}{3}(m_{K}^{2}-m_{\pi }^{2})A(m_{K},m_{\eta
},m_{\pi })
\nonumber \\
&&  \nonumber \\
&&-\sqrt{2}(m_{K}^{2}+m_{\pi }^{2})B(m_{\pi },m_{\pi },m_{K})-6\sqrt{2}%
m_{K}^{2}B(m_{K},m_{K},m_{K})  \nonumber \\
&&-\frac{\sqrt{2}}{2}(m_{K}^{2}+m_{\pi }^{2})B(m_{\pi },m_{K},m_{\pi })-%
\frac{\sqrt{2}}{2}(3m_{K}^{2}-m_{\pi }^{2})B(m_{\eta
},m_{K},m_{\eta })
\nonumber \\
&&+\frac{\sqrt{2}}{6}(m_{K}^{2}-m_{\pi }^{2})B(m_{\pi },m_{K},m_{\eta })-%
\frac{2\sqrt{2}}{3}(m_{K}^{2}-m_{\pi }^{2})B(m_{K},m_{\pi
},m_{\eta })
\nonumber \\
&&+\frac{\sqrt{2}}{6}(m_{K}^{2}-m_{\pi }^{2})B(m_{\eta },m_{K},m_{\pi })+%
\frac{\sqrt{2}}{3}(m_{K}^{2}-m_{\pi }^{2})B(m_{\eta },m_{\pi
},m_{K})
\nonumber \\
&&  \nonumber \\
&&+\sqrt{2}C(m_{K},m_{K},m_{K})-\frac{\sqrt{2}}{2}C(m_{\pi },m_{\pi },m_{K})-%
\frac{\sqrt{2}}{2}C(m_{K},m_{\pi },m_{\pi })  \nonumber \\
&&-\frac{3\sqrt{2}}{2}C(m_{K},m_{\eta },m_{\eta })-\frac{\sqrt{2}}{2}%
C(m_{\pi },m_{\eta },m_{K})-\sqrt{2}C(m_{K},m_{\eta },m_{\pi })  \nonumber \\
&&  \nonumber \\
&&-7\sqrt{2}D(m_{K},m_{K},m_{K})-\frac{5\sqrt{2}}{2}D(m_{\pi
},m_{\pi
},m_{K})-\frac{7\sqrt{2}}{4}D(m_{\pi },m_{K},m_{\pi })  \nonumber \\
&&-\frac{\sqrt{2}}{2}D(m_{\eta },m_{\pi },m_{K})
-\frac{\sqrt{2}}{4}D(m_{\pi
},m_{K},m_{\eta })-2\sqrt{2}D(m_{K},m_{\pi },m_{\eta })  \nonumber \\
&&-\frac{\sqrt{2}}{4}D(m_{\eta },m_{K},m_{\pi })-\frac{3\sqrt{2}}{4}%
D(m_{\eta },m_{K},m_{\eta })  \nonumber \\
&&  \nonumber \\
&&-7\sqrt{2}E(m_{K},m_{K},m_{K})-\frac{5\sqrt{2}}{2}E(m_{\pi
},m_{K},m_{\pi
})-\frac{7\sqrt{2}}{4}E(m_{K},m_{\pi },m_{\pi })  \nonumber \\
&&-\frac{\sqrt{2}}{2}E(m_{\pi },m_{K},m_{\eta }) -\frac{\sqrt{2}}{4}%
E(m_{K},m_{\eta },m_{\pi })-\frac{3\sqrt{2}}{4}E(m_{K},m_{\eta
},m_{\eta })
\nonumber \\
&&-\frac{\sqrt{2}}{4}E(m_{K},m_{\pi },m_{\eta })-2\sqrt{2}E(m_{\pi
},m_{\eta
},m_{K})  \nonumber \\
&&  \nonumber \\
&&+\sqrt{2}F(m_{K},m_{K},m_{K})-\frac{\sqrt{2}}{2}F(m_{\pi },m_{K},m_{\pi })-%
\frac{\sqrt{2}}{2}F(m_{\pi },m_{\pi },m_{K})  \nonumber \\
&&-\frac{3\sqrt{2}}{2}F(m_{\eta },m_{\eta },m_{K}) -\frac{\sqrt{2}}{2}%
F(m_{\eta },m_{K},m_{\pi })-\sqrt{2}F(m_{\eta },m_{\pi },m_{K})  \nonumber \\
&&  \nonumber \\
&&-\sqrt{2}G(m_{K},m_{K},m_{K})-\sqrt{2}G(m_{\pi },m_{K},m_{\pi })+\frac{%
\sqrt{2}}{2}G(m_{K},m_{\pi },m_{\pi })  \nonumber \\
&&+\frac{3\sqrt{2}}{2}G(m_{K},m_{\eta },m_{\eta }) +\frac{\sqrt{2}}{2}%
G(m_{K},m_{\eta },m_{\pi })+\frac{\sqrt{2}}{2}G(m_{K},m_{\pi
},m_{\eta })
\nonumber \\
&&+\frac{3\sqrt{2}}{2}G(m_{\pi },m_{\pi },m_{K})+\sqrt{2}G(m_{\pi
},m_{K},m_{\eta }) -\frac{\sqrt{2}}{2}G(m_{\pi },m_{\eta },m_{K})
\nonumber
\\
&&  \nonumber \\
&&-\sqrt{2}H(m_{K},m_{K},m_{K})+\frac{3\sqrt{2}}{2}H(m_{\pi
},m_{\pi
},m_{K})-\sqrt{2}H(m_{K},m_{\pi },m_{\pi })  \nonumber \\
&&+\frac{\sqrt{2}}{2}H(m_{\pi },m_{K},m_{\pi })
+\sqrt{2}H(m_{K},m_{\pi
},m_{\eta })+\frac{\sqrt{2}}{2}H(m_{\pi },m_{K},m_{\eta })  \nonumber \\
&&+\frac{3\sqrt{2}}{2}H(m_{\eta },m_{K},m_{\eta })+\frac{\sqrt{2}}{2}%
H(m_{\eta },m_{K},m_{\pi }) -\frac{\sqrt{2}}{2}H(m_{\eta },m_{\pi
},m_{K})
\nonumber \\
&&  \nonumber \\
&&+8\sqrt{2}I(m_{K},m_{K},m_{K})+\sqrt{2}I(m_{K},m_{\pi },m_{\pi })+\sqrt{2}%
I(m_{\pi },m_{K},m_{\pi })  \nonumber \\
&&+\frac{5\sqrt{2}}{4}I(m_{\pi },m_{\pi },m_{K}) -\frac{3\sqrt{2}}{4}%
I(m_{\eta },m_{\eta },m_{K})+\sqrt{2}I(m_{K},m_{\eta },m_{\pi })
\nonumber
\\
&&+\sqrt{2}I(m_{\eta },m_{K},m_{\pi })-\frac{\sqrt{2}}{4}I(m_{\pi
},m_{\eta
},m_{K})-\frac{\sqrt{2}}{4}I(m_{\eta },m_{\pi },m_{K})  \nonumber \\
&&  \nonumber \\
&&-\sqrt{2}J(m_{K},m_{K},m_{K})-\sqrt{2}J(m_{\pi },m_{\pi },m_{K})+\frac{%
\sqrt{2}}{2}J(m_{K},m_{\pi },m_{\pi })  \nonumber \\
&&-\frac{\sqrt{2}}{2}J(m_{\pi },m_{K},m_{\eta }) +\frac{3\sqrt{2}}{2}%
J(m_{\pi },m_{K},m_{\pi })+\frac{\sqrt{2}}{2}J(m_{K},m_{\pi
},m_{\eta })
\nonumber \\
&&+\frac{\sqrt{2}}{2}J(m_{K},m_{\eta },m_{\pi })+\frac{3\sqrt{2}}{2}%
J(m_{K},m_{\eta },m_{\eta }) +\sqrt{2}J(m_{\pi },m_{\eta },m_{K})
\nonumber
\\
&&  \nonumber \\
&&+8\sqrt{2}K(m_{K},m_{K},m_{K})+\sqrt{2}K(m_{K},m_{\pi },m_{\pi })+\sqrt{2}%
K(m_{\pi },m_{\pi },m_{K})  \nonumber \\
&&+\frac{5\sqrt{2}}{4}K(m_{\pi },m_{K},m_{\pi }) -\frac{3\sqrt{2}}{4}%
K(m_{\eta },m_{K},m_{\eta })-\frac{\sqrt{2}}{4}K(m_{\pi
},m_{K},m_{\eta })
\nonumber \\
&&-\frac{\sqrt{2}}{4}K(m_{\eta },m_{K},m_{\pi })+\sqrt{2}K(m_{\eta
},m_{\pi
},m_{K})+\sqrt{2}K(m_{K},m_{\pi },m_{\eta })  \nonumber \\
&&  \nonumber \\
&&-\sqrt{2}L(m_{K},m_{K},m_{K})-\sqrt{2}L(m_{K},m_{\pi },m_{\pi })+\frac{%
\sqrt{2}}{2}L(m_{\pi },m_{\pi },m_{K})  \nonumber \\
&&+\frac{3\sqrt{2}}{2}L(m_{\eta },m_{\eta },m_{K}) +\frac{\sqrt{2}}{2}%
L(m_{\eta},m_{\pi},m_{K})+\sqrt{2}L(m_{K},m_{\eta },m_{\pi })
\nonumber \\ &&+\frac{3\sqrt{2}}{2}L(m_{\pi },m_{K},m_{\pi
})+\frac{\sqrt{2}}{2}L(m_{\pi },m_{\eta },m_{K})
-\frac{\sqrt{2}}{2}L(m_{\eta },m_{K},m_{\pi })\} %
\\
\nn \\  \nonumber \\ %
 M_{A,\pi,3-point }
&=&\frac{i\sqrt{2}}{12F^{3}}eG_{F} l^{\mu }\varepsilon
^{\nu }\cos \theta\{\sqrt{2}(m_{K}^{2}+m_{\pi }^{2})A(m_{K},m_{\pi },m_{K})%
\nn\\
&&+\frac{20%
\sqrt{2}}{3}m_{\pi }^{2}A(m_{\pi },m_{\pi },m_{\pi })
+\frac{2\sqrt{2}}{3}(m_{K}^{2}%
+m_{\pi }^{2})A(m_{\pi },m_{K},m_{K})%
\nn\\
&&-\frac{%
\sqrt{2}}{3}(m_{\pi }^{2}-m_{K}^{2})A(m_{K},m_{K},m_{\eta })  \nonumber \\
&&  \nonumber \\
&&-\sqrt{2}(m_{K}^{2}+m_{\pi }^{2})B(m_{K},m_{K},m_{\pi })-\frac{2\sqrt{2}}{3%
}(m_{K}^{2}+m_{\pi }^{2})B(m_{K},m_{\pi },m_{K})  \nonumber \\
&&-\frac{20\sqrt{2}}{3}m_{\pi }^{2}B(m_{\pi },m_{\pi },m_{\pi })+\frac{2%
\sqrt{2}}{3}(m_{\pi }^{2}-m_{K}^{2})B(m_{\eta },m_{K},m_{K})  \nonumber \\
&&-\frac{\sqrt{2}}{3}(m_{\pi }^{2}-m_{K}^{2})B(m_{K},m_{K},m_{\eta
})
\nonumber \\
&&  \nonumber \\
&&+\sqrt{2}C(m_{K},m_{\pi },m_{K})-\frac{4\sqrt{2}}{3}C(m_{\pi
},m_{\pi
},m_{\pi })-\frac{2\sqrt{2}}{3}C(m_{\pi },m_{K},m_{K})  \nonumber \\
&&-\frac{3\sqrt{2}}{2}C(m_{K},m_{K},m_{\pi }) -\frac{\sqrt{2}}{2}%
C(m_{K},m_{K},m_{\eta })  \nonumber \\
&&  \nonumber \\
&&-\frac{5\sqrt{2}}{2}D(m_{K},m_{K},m_{\pi})-\frac{20\sqrt{2}}{3}D(m_{\pi
},m_{\pi },m_{\pi })%
\nn\\ &&-\frac{10\sqrt{2}}{3}D(m_{K},m_{\pi },m_{K})
-\frac{\sqrt{2}}{2}D(m_{K},m_{K},m_{\eta }) -2\sqrt{2}D(m_{\eta
},m_{K},m_{K})  \nonumber \\ &&  \nonumber \\
&&-\frac{5\sqrt{2}}{2}E(m_{K},m_{\pi},m_{K})-\frac{20\sqrt{2}}{3}E(m_{\pi
},m_{\pi },m_{\pi })%
\nn\\ &&-\frac{10\sqrt{2}}{3}E(m_{\pi },m_{K},m_{K})
-2\sqrt{2}E(m_{K},m_{K},m_{\eta
})-\frac{\sqrt{2}}{2}E(m_{K},m_{\eta },m_{K})  \nonumber \\ &&
\nonumber \\
&&-\frac{\sqrt{2}}{2}F(m_{\pi},m_{K},m_{K})-\frac{4\sqrt{2}}{3}F(m_{\pi
},m_{\pi },m_{\pi })-\frac{2\sqrt{2}}{3}F(m_{K},m_{K},m_{\pi })
\nonumber \\ &&-\frac{\sqrt{2}}{2}F(m_{\eta },m_{K},m_{K})
\nonumber \\ &&  \nonumber \\ &&-\sqrt{2}G(m_{K},m_{\pi
},m_{K})+\frac{4\sqrt{2}}{3}G(m_{\pi },m_{\pi },m_{\pi
})+\frac{2\sqrt{2}}{3}G(m_{\pi },m_{K},m_{K})  \nonumber \\
&&+\frac{3\sqrt{2}}{2}G(m_{K},m_{K},m_{\pi })-\frac{\sqrt{2}}{2}%
G(m_{K},m_{K},m_{\eta })+\sqrt{2}G(m_{K},m_{\eta },m_{K})  \nonumber \\
&&  \nonumber \\
&&-\sqrt{2}H(m_{\pi},m_{K},m_{K})+\frac{2\sqrt{2}}{3}H(m_{K},m_{\pi },m_{K})+%
\frac{4\sqrt{2}}{3}H(m_{\pi },m_{\pi },m_{\pi })  \nonumber \\
&&+\frac{3\sqrt{2}}{2}H(m_{K},m_{K},m_{\pi }) +\sqrt{2}H(m_{\eta
},m_{K},m_{K})-\frac{\sqrt{2}}{2}H(m_{K},m_{K},m_{\eta })  \nonumber \\
&&  \nonumber \\
&&+\sqrt{2}I(m_{\pi },m_{K},m_{K})+\sqrt{2}I(m_{K},m_{\pi },m_{K})+\frac{16%
\sqrt{2}}{3}I(m_{\pi },m_{\pi },m_{\pi })  \nonumber \\
&&+\frac{8\sqrt{2}}{3}I(m_{K},m_{K},m_{\pi
})+\sqrt{2}I(m_{K},m_{\eta
},m_{K})+\sqrt{2}I(m_{\eta },m_{K},m_{K})  \nonumber \\
&&  \nonumber \\
&&-\sqrt{2}J(m_{K},m_{K},m_{\pi })+\frac{4\sqrt{2}}{3}J(m_{\pi
},m_{\pi
},m_{\pi })+\frac{2\sqrt{2}}{3}J(m_{\pi },m_{K},m_{K})  \nonumber \\
&&+\frac{3\sqrt{2}}{2}J(m_{K},m_{\pi },m_{K})
+\sqrt{2}J(m_{K},m_{K},m_{\eta
})-\frac{\sqrt{2}}{2}J(m_{K},m_{\eta },m_{K})  \nonumber \\
&&  \nonumber \\
&&+\sqrt{2}K(m_{K},m_{K},m_{\pi })+\sqrt{2}K(m_{\pi },m_{K},m_{K})+\frac{16%
\sqrt{2}}{3}K(m_{\pi },m_{\pi },m_{\pi })  \nonumber \\
&&+\frac{8\sqrt{2}}{3}K(m_{K},m_{\pi },m_{K}) +\sqrt{2}K(m_{\eta
},m_{K},m_{K})+\sqrt{2}K(m_{K},m_{K},m_{\eta })  \nonumber \\
&&  \nonumber \\
&&-\sqrt{2}L(m_{\pi },m_{K},m_{K})+\frac{4\sqrt{2}}{3}L(m_{\pi
},m_{\pi
},m_{\pi })+\frac{2\sqrt{2}}{3}L(m_{K},m_{K},m_{\pi })  \nonumber \\
&&+\frac{3\sqrt{2}}{2}L(m_{K},m_{\pi },m_{K})-\frac{\sqrt{2}}{2}%
L(m_{K},m_{\eta },m_{K})+\sqrt{2}L(m_{\eta },m_{K},m_{K})\}\,.
\end{eqnarray}

\newpage
\section*{Appendix B}

We list the functions $\{II, III, \cdots,A,B,\cdots,L\}$ in
Figures \ref{fig3} and \ref {fig4} by scalar integrals of
$P_{\alpha 1,\alpha 2,\alpha 3}^{ab}$ and one-loop tadpole
integrals $T_{1}$, $T_{2}$ in the Euclidian space. We note that
the function $I$ does not contain $g^{\mu\nu}$ and thus it has no
contribution to $F_A$. For simplicity, we only give the formulas
related to terms with $g_{\mu \nu }$ due to the definition of
$F_{A}$. We have

\begin{eqnarray}
II(m_{1},m_{2},m_{3})&=&
{-ig^{\mu\nu}\over (2\pi )^{2n}(n-1)}
\left[T_{1}(m_{2}^{2})T_{1}(m_{3}^{2})-m_{1}^{2}P_{111}^{00}(m_{1},m_{2},m_{3})%
\right.%
 \nonumber \\ &&%
 \left. -\frac{1}{\ell
^{2}}P_{111}^{20}(m_{1},m_{2},m_{3})\right]+\cdots\,,
\\ \nonumber \\ %
III(m_{1},m_{2},m_{3})&=&
{-ig^{\mu\nu}\over (2\pi )^{2n}(n-1)}
\left[\frac{1}{2}T_{1}(m_{1}^{2})T_{1}(m_{2}^{2})-\frac{1}{2}%
T_{1}(m_{2}^{2})T_{1}(m_{3}^{2})-\frac{1}{2}T_{1}(m_{1}^{2})T_{1}(m_{3}^{2})%
\right. \nonumber \\ %
&&+\frac{1}{2}(m_{1}^{2}+m_{2}^{2}-m_{3}^{2}-\ell
^{2})P_{111}^{00}(m_{1},m_{2},m_{3})-P_{111}^{10}(m_{1},m_{2},m_{3})
\nonumber \\ %
&&\left.-P_{111}^{01}(m_{1},m_{2},m_{3})-\frac{1}{\ell ^{2}}%
P_{111}^{11}(m_{1},m_{2},m_{3})\right]+\cdots\,,
\\ \nonumber \\
A(m_{1},m_{2},m_{3})&=&\frac{-2g^{\mu \nu }}{(2\pi
)^{2n}(n-1)}\int
dx(-1)\left[P_{111}^{00}(m_{1},m_{2},m_{3})-m_{1}^{2}P_{211}^{00}(m_{1},m_{2},m_{3})
\right.\nonumber \\ &&\left. -\frac{1}{\ell ^{\prime
2}}P_{211}^{20}(m_{1},m_{2},m_{3})\right] +\cdots\,,
\\ \nonumber
\\ %
B(m_{1},m_{2},m_{3})&=&\frac{-g^{\mu \nu }}{(2\pi )^{2n}(n-1)}\int
dx(-1)\left[T_{2}(m_{2}^{2})T_{1}(m_{1}^{2})-T_{2}(m_{2}^{2})T_{1}(m_{3}^{2})
\right.\nonumber \\ &&-P_{111}^{00}(m_{2},m_{1},m_{3})
+(m_{1}^{2}+m_{2}^{2}-m_{3}^{2}-\ell ^{\prime
2})P_{211}^{00}(m_{2},m_{1},m_{3})  \nonumber \\
&&-2P_{211}^{10}(m_{2},m_{1},m_{3}) -2P_{211}^{01}(m_{2},m_{1},m_{3})%
\nn\\%
&&\left.-\frac{2%
}{\ell ^{\prime
2}}P_{211}^{11}(m_{2},m_{1},m_{3})\right]+\cdots\,,
\end{eqnarray}
\begin{eqnarray}
&&C(m_{1},m_{2},m_{3})+F(m_{3},m_{1},m_{2})  \nonumber \\ &=&
{g^{\mu \nu }\over (2\pi )^{2n}(n-1)} \int
dx\left[2(1-x)(p_{M}\cdot k)P_{111}^{00}(m_{1},m_{2},m_{3})%
\right. \nonumber \\
&&-\frac{2(p_{M}\cdot \ell ^{\prime })}{\ell ^{\prime 2}}%
P_{111}^{10}(m_{1},m_{2},m_{3}) -T_{1}(m_{1}^{2})T_{1}(m_{2}^{2})
+T_{1}(m_{2}^{2})T_{1}(m_{3}^{2})-T_{1}(m_{1}^{2})T_{1}(m_{3}^{2})  \nonumber
\\
&&+2P_{111}^{10}(m_{1},m_{2},m_{3}) +(-m_{1}^{2}+m_{2}^{2}+m_{3}^{2}+\ell
^{\prime 2})P_{111}^{00}(m_{1},m_{2},m_{3})  \nonumber \\
&& -2m_{1}^{2}(1-x)(p_{M}\cdot k)P_{211}^{00}(m_{1},m_{2},m_{3})+2m_{1}^{2}%
\frac{(p_{M}\cdot \ell ^{\prime })}{\ell ^{\prime 2}}%
P_{211}^{10}(m_{1},m_{2},m_{3})  \nonumber \\
&&+m_{1}^{2}T_{2}(m_{1}^{2})T_{1}(m_{2}^{2})
-m_{1}^{2}P_{111}^{00}(m_{1},m_{2},m_{3})+m_{1}^{2}T_{2}(m_{1}^{2})T_{1}(m_{3}^{2})
\nonumber \\
&&-2m_{1}^{2}P_{211}^{10}(m_{1},m_{2},m_{3})-m_{1}^{2}(-m_{1}^{2}
+m_{2}^{2}+m_{3}^{2}+\ell ^{\prime 2})P_{211}^{00}(m_{1},m_{2},m_{3})
\nonumber \\
&& -\frac{2(1-x)(p_{M}\cdot k)}{\ell ^{\prime 2}}%
P_{211}^{20}(m_{1},m_{2},m_{3})+\frac{2(p_{M}\cdot \ell ^{\prime })}{\ell
^{\prime 4}}P_{211}^{30}(m_{1},m_{2},m_{3})  \nonumber \\
&&-{\frac{2}{\ell^{\prime 2}}}P_{211}^{30}(m_{1},m_{2},m_3)-\frac{1}{\ell
^{\prime 2}}(-m_{1}^{2}+m_{2}^{2}+m_{3}^{2}+\ell ^{\prime
2})P_{211}^{20}(m_{1},m_{2},m_{3})  \nonumber \\
&& +\frac{1}{\ell ^{\prime 2}}T_{1}(m_{2}^{2})\int \frac{d^{n}p(p\cdot \ell
^{\prime })^{2}}{(p^{2}+m_{1}^{2})^{2}} -\frac{1}{\ell ^{\prime 2}}%
P_{111}^{20}(m_{1},m_{2},m_{3})%
\nn\\
&&\left.+\frac{1}{\ell ^{\prime 2}}%
T_{1}(m_{3}^{2})\int \frac{d^{n}p(p\cdot \ell ^{\prime })^{2}}{%
(p^{2}+m_{1}^{2})^{2}}\right]+\cdots\,,
\end{eqnarray}
\begin{eqnarray}
&&D(m_{2},m_{1},m_{3})+E(m_{1},m_{3},m_{2})  \nonumber \\
&=&{-g^{\mu \nu }\over (2\pi )^{2n}(n-1)}\int dx \left\{%
T_{1}(m_{1}^{2}) \int d^{n}p\frac{p^2}{(p^2+m_2^2)^2} -(1-x)(\ell
^{\prime }\cdot k)T_{2}(m_{2}^{2})T_{1}(m_{1}^{2}) %
\right.\nonumber \\
&& -\frac{(p_{M}\cdot \ell ^{\prime })}{\ell ^{\prime 2}}%
P_{111}^{01}(m_{2},m_{1},m_{3})+(1-x)(\ell ^{\prime }\cdot
k)P_{111}^{00}(m_{2},m_{1},m_{3})  \nonumber \\
&&+\frac{(1-x)(\ell ^{\prime }\cdot k)}{\ell ^{\prime 2}}%
P_{111}^{01}(m_{2},m_{1},m_{3})+\frac{(1-x)(\ell ^{\prime }\cdot k)}{\ell
^{\prime 2}}P_{111}^{10}(m_{2},m_{1},m_{3})  \nonumber \\
&&-\frac{1}{2}T_{1}(m_{1}^{2})T_{1}(m_{3}^{2}) +\frac{1}{2}%
m_{2}^{2}P_{111}^{00}(m_{2},m_{1},m_{3})-\frac{1}{2}%
T_{1}(m_{2}^{2})T_{1}(m_{1}^{2})+\frac{1}{2}T_{1}(m_{2}^{2})T_{1}(m_{3}^{2})
\nonumber \\
&& -\frac{1}{2}(m_{1}^{2}-\ell ^{\prime
2}-m_{3}^{2})P_{111}^{00}(m_{2},m_{1},m_{3})+P_{111}^{01}(m_{2},m_{1},m_{3})
+(p_{M}\cdot \ell ^{\prime })T_{2}(m_{2}^{2})T_{1}(m_{3}^{2})
\nonumber \\ && +(m_{1}^{2}+m_{2}^{2}-\ell ^{\prime 2}-m_{3}^{2})
\left[ \frac{(p_{M}\cdot
\ell ^{\prime })}{\ell ^{\prime 2}}P_{211}^{01}(m_{2},m_{1},m_{3})-(1-x)(%
\ell ^{\prime }\cdot k)P_{211}^{00}(m_{2},m_{1},m_{3})\right.
\nonumber
\\
&& -\frac{(1-x)(\ell ^{\prime }\cdot k)}{\ell ^{\prime 2}}%
P_{211}^{01}(m_{2},m_{1},m_{3})-\frac{(1-x)(\ell ^{\prime }\cdot k)}{\ell
^{\prime 2}}P_{211}^{10}(m_{2},m_{1},m_{3})  \nonumber \\
&&+\frac{1}{2}P_{111}^{00}(m_{2},m_{1},m_{3}) -\frac{1}{2}%
m_{2}^{2}P_{211}^{00}(m_{2},m_{1},m_{3})+\frac{1}{2}%
T_{2}(m_{2}^{2})T_{1}(m_{1}^{2})-\frac{1}{2}T_{2}(m_{2}^{2})T_{1}(m_{3}^{2})
\nonumber \\ %
&& \left.+\frac{1}{2}(m_{1}^{2}-\ell ^{\prime
2}-m_{3}^{2})P_{211}^{00}(m_{2},m_{1},m_{3})-P_{211}^{01}(m_{2},m_{1},m_{3})\right]
\nonumber \\
&& -2\frac{(p_{M}\cdot \ell ^{\prime })}{\ell ^{\prime 2}}%
P_{211}^{02}(m_{2},m_{1},m_{3})+2(1-x)(\ell ^{\prime }\cdot
k)P_{211}^{01}(m_{2},m_{1},m_{3})  \nonumber \\
&& +\frac{2(1-x)(\ell ^{\prime }\cdot k)}{\ell ^{\prime 2}}%
P_{211}^{02}(m_{2},m_{1},m_{3})+\frac{2(1-x)(\ell ^{\prime }\cdot k)}{\ell
^{\prime 2}}P_{211}^{11}(m_{2},m_{1},m_{3})  \nonumber \\
&&-P_{111}^{01}(m_{2},m_{1},m_{3})
+m_{2}^{2}P_{211}^{01}(m_{2},m_{1},m_{3})-\ell ^{\prime
2}T_{2}(m_{2}^{2})T_{1}(m_{3}^{2})  \nonumber \\
&&-(m_{1}^{2}-\ell ^{\prime 2}-m_{3}^{2})P_{211}^{01}(m_{2},m_{1},m_{3})
+2P_{211}^{02}(m_{2},m_{1},m_{3})  \nonumber \\
&& -2\frac{(p_{M}\cdot \ell ^{\prime })}{\ell ^{\prime 2}}%
P_{211}^{11}(m_{2},m_{1},m_{3})+2(1-x)(\ell ^{\prime }\cdot
k)P_{211}^{10}(m_{2},m_{1},m_{3})  \nonumber \\
&& +\frac{2(1-x)(\ell ^{\prime }\cdot k)}{\ell ^{\prime 2}}%
P_{211}^{11}(m_{2},m_{1},m_{3})+\frac{2(1-x)(\ell ^{\prime }\cdot k)}{\ell
^{\prime 2}}P_{211}^{20}(m_{2},m_{1},m_{3})  \nonumber \\
&&-P_{111}^{10}(m_{2},m_{1},m_{3})
+m_{2}^{2}P_{211}^{10}(m_{2},m_{1},m_{3})-(m_{1}^{2}-\ell ^{\prime
2}-m_{3}^{2})P_{211}^{10}(m_{2},m_{1},m_{3})  \nonumber \\
&&+2P_{211}^{11}(m_{2},m_{1},m_{3}) -\frac{2(p_{M}\cdot \ell ^{\prime })}{%
\ell ^{\prime 4}}P_{211}^{12}(m_{2},m_{1},m_{3})  \nonumber \\
&&+\frac{2(1-x)(\ell ^{\prime }\cdot k)}{\ell ^{\prime 2}}%
P_{211}^{11}(m_{2},m_{1},m_{3}) +2\frac{(1-x)(\ell ^{\prime }\cdot k)}{\ell
^{\prime 4}}P_{211}^{12}(m_{2},m_{1},m_{3})  \nonumber \\
&&+\frac{2(1-x)(\ell ^{\prime }\cdot k)}{\ell ^{\prime 4}}%
P_{211}^{21}(m_{2},m_{1},m_{3})-\frac{1}{\ell ^{\prime 2}}%
P_{111}^{11}(m_{2},m_{1},m_{3})  \nonumber \\
&& +\frac{1}{\ell ^{\prime 2}}m_{2}^{2}P_{211}^{11}(m_{2},m_{1},m_{3})-\frac{%
(m_{1}^{2}-\ell ^{\prime 2}-m_{3}^{2})}{\ell ^{\prime 2}}%
P_{211}^{11}(m_{2},m_{1},m_{3})  \nonumber \\ &&
\left.+\frac{2}{\ell ^{\prime 2}}P_{211}^{12}(m_{2},m_{1},m_{3})
-\frac{1}{\ell
^{\prime 2}}T_{1}(m_{3}^{2})\int \frac{d^{n}p(p\cdot \ell ^{\prime })^{2}}{%
(p^{2}+m_{2}^{2})^{2}}\right\}+\cdots\,,
\end{eqnarray}

\begin{eqnarray}
&&H(m_{2},m_{1},m_{3})+J(m_{1},m_{3},m_{2})  \nonumber \\
&=&{-g^{\mu \nu }\over (2\pi )^{2n}(n-1)}\int dx\left\{\frac{%
2(p_{M}\cdot \ell ^{\prime })}{\ell ^{\prime 2}}%
P_{111}^{01}(m_{2},m_{1},m_{3})+2P_{111}^{10}(m_{2},m_{1},m_{3}) %
\right. \nonumber
\\
&&-2(1-x)(\ell ^{\prime }\cdot k)P_{111}^{00}(m_{2},m_{1},m_{3})-\frac{%
2(1-x)(\ell ^{\prime }\cdot k)}{\ell ^{\prime 2}}%
P_{111}^{10}(m_{2},m_{1},m_{3})  \nonumber \\
&&-\frac{2(1-x)(\ell ^{\prime }\cdot k)}{\ell ^{\prime 2}}%
P_{111}^{01}(m_{2},m_{1},m_{3})+T_{1}(m_{2}^{2})T_{1}(m_{1}^{2})  \nonumber
\\
&&+(-m_{2}^{2}+m_{1}^{2}-\ell ^{\prime
2}-m_{3}^{2})P_{111}^{00}(m_{2},m_{1},m_{3})-2P_{111}^{10}(m_{2},m_{1},m_{3})
\nonumber \\
&&-2P_{111}^{01}(m_{2},m_{1},m_{3})+T_{1}(m_{1}^{2})T_{1}(m_{3}^{2})
-T_{1}(m_{2}^{2})T_{1}(m_{3}^{2})  \nonumber \\
&&-{\frac{2m_{2}^{2}(p_{M}\cdot \ell ^{\prime })}{\ell ^{\prime 2}}}%
P_{211}^{01}(m_{2},m_{1},m_{3})-2m_{2}^{2}P_{211}^{10}(m_{2},m_{1},m_{3})
\nonumber \\
&&+2m_{2}^{2}(1-x)(\ell ^{\prime }\cdot k)P_{211}^{00}(m_{2},m_{1},m_{3})+%
\frac{2m_{2}^{2}(1-x)(\ell ^{\prime }\cdot k)}{\ell ^{\prime 2}}%
P_{211}^{10}(m_{2},m_{1},m_{3})  \nonumber \\
&&+\frac{2m_{2}^{2}(1-x)(\ell ^{\prime }\cdot k)}{\ell ^{\prime 2}}%
P_{211}^{01}(m_{2},m_{1},m_{3})-m_{2}^{2}T_{1}(m_{2}^{2})T_{1}(m_{1}^{2})
\nonumber \\
&&-m_{2}^{2}(-m_{2}^{2}+m_{1}^{2}-\ell ^{\prime
2}-m_{3}^{2})P_{211}^{00}(m_{2},m_{1},m_{3})+2m_{2}^{2}P_{211}^{10}(m_{2},m_{1},m_{3})
\nonumber \\
&&+2m_{2}^{2}P_{211}^{01}(m_{2},m_{1},m_{3})
+m_{2}^{2}T_{2}(m_{2}^{2})T_{1}(m_{3}^{2})-m_{2}^{2}P_{111}^{00}(m_{2},m_{1},m_{3})
\nonumber \\
&&-\frac{2(p_{M}\cdot \ell ^{\prime })}{\ell ^{\prime 4}}%
P_{211}^{21}(m_{2},m_{1},m_{3})-\frac{2}{\ell ^{\prime 2}}%
P_{211}^{30}(m_{2},m_{1},m_{3})  \nonumber \\
&&+\frac{2(1-x)(\ell ^{\prime }\cdot k)}{\ell ^{\prime 2}}%
P_{211}^{20}(m_{2},m_{1},m_{3}) +2\frac{(1-x)(\ell ^{\prime }\cdot k)}{\ell
^{\prime 4}}P_{211}^{30}(m_{2},m_{1},m_{3})  \nonumber \\
&&+\frac{2(1-x)(\ell ^{\prime }\cdot k)}{\ell ^{\prime 4}}%
P_{211}^{21}(m_{2},m_{1},m_{3})- T_{1}(m_{1}^{2})\int
\frac{d^{n}p\frac{(p\cdot \ell ^{\prime })^{2}}{\ell ^{\prime
2}}}{(p^{2}+m_{2}^{2})^{2}} \nonumber \\
&&-\frac{(-m_{2}^{2}+m_{1}^{2}-\ell ^{\prime 2}-m_{3}^{2})}{\ell ^{\prime 2}}%
P_{211}^{20}(m_{2},m_{1},m_{3})+\frac{2}{\ell ^{\prime 2}}%
P_{211}^{30}(m_{2},m_{1},m_{3})  \nonumber \\
&&\left.+\frac{2}{\ell ^{\prime 2}}P_{211}^{21}(m_{2},m_{1},m_{3})
-\frac{1}{\ell ^{\prime
2}}P_{111}^{20}(m_{2},m_{1},m_{3})+T_{1}(m_{3}^{2})\int
\frac{d^{n}p\frac{(p\cdot \ell ^{\prime })^{2}}{\ell ^{\prime
2}}}{(p^{2}+m_{2}^{2})^{2}} \right\}%
\nn\\ &&+\cdots\,,
\end{eqnarray}
%
\begin{eqnarray}
&&I(m_{3},m_{2},m_{1})+K(m_{2},m_{1},m_{3})  \nonumber \\
&=&{g^{\mu \nu }\over (2\pi )^{2n}(n-1)}\int dx\left\{(p_{M}\cdot
\ell ^{\prime })T_{2}(m_{3}^{2})T_{1}(m_{2}^{2}) \right.\nonumber
\\
&&-(p_{M}\cdot \ell ^{\prime })P_{111}^{00}(m_{3},m_{2},m_{1})-\frac{%
(p_{M}\cdot \ell ^{\prime })}{\ell ^{\prime 2}}%
P_{111}^{10}(m_{3},m_{2},m_{1})  \nonumber \\
&& -\frac{(p_{M}\cdot \ell ^{\prime })}{\ell ^{\prime 2}}%
P_{111}^{01}(m_{3},m_{2},m_{1}) +\frac{(1-x)(\ell ^{\prime }\cdot k)}{\ell
^{\prime 2}}P_{111}^{01}(m_{3},m_{2},m_{1})  \nonumber \\
&&-\frac{1}{2}T_{1}(m_{3}^{2})T_{1}(m_{2}^{2})+\frac{1}{2}%
T_{1}(m_{2}^{2})T_{1}(m_{1}^{2})+\frac{1}{2}T_{1}(m_{3}^{2})T_{1}(m_{1}^{2})
\nonumber \\
&&-\frac{1}{2}(m_{2}^{2}-m_{1}^{2}-\ell ^{\prime
2}+m_{3}^{2})P_{111}^{00}(m_{3},m_{2},m_{1}) +P_{111}^{10}(m_{3},m_{2},m_{1})
\nonumber \\
&&+P_{111}^{01}(m_{3},m_{2},m_{1}) -(1-x)(\ell ^{\prime }\cdot
k)T_{2}(m_{3}^{2})T_{1}(m_{1}^{2})+T_{1}(m_{1}^{2})\int \frac{d^{n}p(p)^{2}}{%
(p^{2}+m_{3}^{2})^{2}}  \nonumber \\
&&+(m_{3}^{2}+m_{2}^{2}-\ell ^{\prime 2}-m_{1}^{2})\times \lbrack
(p_{M}\cdot \ell ^{\prime })P_{211}^{00}(m_{3},m_{2},m_{1})  \nonumber \\
&&+\frac{(p_{M}\cdot \ell ^{\prime })}{\ell ^{\prime 2}}%
P_{211}^{10}(m_{3},m_{2},m_{1})+\frac{(p_{M}\cdot \ell ^{\prime })}{\ell
^{\prime 2}}P_{211}^{01}(m_{3},m_{2},m_{1})  \nonumber \\
&&-\frac{(1-x)(\ell ^{\prime }\cdot k)}{\ell ^{\prime 2}}%
P_{211}^{01}(m_{3},m_{2},m_{1})+\frac{1}{2}T_{2}(m_{3}^{2})T_{1}(m_{2}^{2})-%
\frac{1}{2}P_{111}^{00}(m_{3},m_{2},m_{1})  \nonumber \\
&&-\frac{1}{2}T_{2}(m_{3}^{2})T_{1}(m_{1}^{2})+\frac{1}{2}%
(m_{3}^{2}+m_{2}^{2}-\ell ^{\prime
2}-m_{1}^{2})P_{211}^{00}(m_{3},m_{2},m_{1})  \nonumber \\
&&-P_{211}^{10}(m_{3},m_{2},m_{1})-P_{211}^{01}(m_{3},m_{2},m_{1})]
\nonumber \\
&&-2(p_{M}\cdot \ell ^{\prime })P_{211}^{10}(m_{3},m_{2},m_{1})-\frac{%
2(p_{M}\cdot \ell ^{\prime })}{\ell ^{\prime 2}}%
P_{211}^{20}(m_{3},m_{2},m_{1})  \nonumber \\
&&-\frac{2(p_{M}\cdot \ell ^{\prime })}{\ell ^{\prime 2}}%
P_{211}^{11}(m_{3},m_{2},m_{1}) +2\frac{(1-x)(\ell ^{\prime }\cdot k)}{\ell
^{\prime 2}}P_{211}^{11}(m_{3},m_{2},m_{1})  \nonumber \\
&&+P_{111}^{10}(m_{3},m_{2},m_{1}) -(m_{3}^{2}+m_{2}^{2}-\ell ^{\prime
2}-m_{1}^{2})P_{211}^{10}(m_{3},m_{2},m_{1})
+2P_{211}^{20}(m_{3},m_{2},m_{1})  \nonumber \\
&&+2P_{211}^{11}(m_{3},m_{2},m_{1}) -2(p_{M}\cdot \ell ^{\prime
})P_{211}^{01}(m_{3},m_{2},m_{1})-\frac{2(p_{M}\cdot \ell ^{\prime })}{\ell
^{\prime 2}}P_{211}^{11}(m_{3},m_{2},m_{1})  \nonumber \\
&&-\frac{2(p_{M}\cdot \ell ^{\prime })}{\ell ^{\prime 2}}%
P_{211}^{02}(m_{3},m_{2},m_{1}) +2\frac{(1-x)(\ell ^{\prime }\cdot k)}{\ell
^{\prime 2}}P_{211}^{02}(m_{3},m_{2},m_{1})+P_{111}^{01}(m_{3},m_{2},m_{1})
\nonumber \\
&&-\ell ^{\prime 2}T_{2}(m_{3}^{2})T_{1}(m_{1}^{2})
-(m_{3}^{2}+m_{2}^{2}-\ell ^{\prime
2}-m_{1}^{2})P_{211}^{01}(m_{3},m_{2},m_{1})  \nonumber \\
&&+2P_{211}^{11}(m_{3},m_{2},m_{1})+2P_{211}^{02}(m_{3},m_{2},m_{1}) -\frac{%
2(p_{M}\cdot \ell ^{\prime })}{\ell ^{\prime 2}}%
P_{211}^{11}(m_{3},m_{2},m_{1})  \nonumber \\
&&-2\frac{(p_{M}\cdot \ell ^{\prime })}{\ell ^{\prime 4}}%
P_{211}^{21}(m_{3},m_{2},m_{1})-2\frac{(p_{M}\cdot \ell ^{\prime })}{\ell
^{\prime 4}}P_{211}^{12}(m_{3},m_{2},m_{1})  \nonumber \\
&&+2\frac{(1-x)(\ell ^{\prime }\cdot k)}{\ell ^{\prime 4}}%
P_{211}^{12}(m_{3},m_{2},m_{1})+\frac{1}{\ell ^{\prime 2}}%
P_{111}^{11}(m_{3},m_{2},m_{1})  \nonumber \\
&&-\frac{(m_{3}^{2}+m_{2}^{2}-\ell ^{\prime 2}-m_{1}^{2})}{\ell ^{\prime 2}}%
P_{211}^{11}(m_{3},m_{2},m_{1})+\frac{2}{\ell ^{\prime 2}}%
P_{211}^{21}(m_{3},m_{2},m_{1})
  \nonumber \\
 &&\left.+\frac{2}{\ell^{\prime 2}}P_{211}^{12}(m_{3},m_{2},m_{1})
-\frac{1}{\ell^{\prime 2}}T_{1}(m_{1}^{2})
\int \frac{d^{n}p(p\cdot \ell ^{\prime })^2}{%
(p^2+m_3^2)^2}\right\}+\cdots\,,
\end{eqnarray}
\begin{eqnarray}
&&G(m_{1},m_{2},m_{3})+L(m_{3},m_{1},m_{2})  \nonumber \\
&=&{g^{\mu \nu }\over (2\pi )^{2n}(n-1)}\int dx\left\{(p_{M}\cdot
k)(1-x)T_{2}(m_{1}^{2})T_{1}(m_{2}^{2})-T_{2}(m_{1}^{2})\int \frac{%
d^{n}p\,p^{2}}{(p^{2}+m_{2}^{2})}  \right.\nonumber \\
&&+\frac{(p_{M}\cdot \ell ^{\prime })}{\ell ^{\prime 2}}%
P_{111}^{10}(m_{1},m_{2},m_{3})-(p_{M}\cdot
k)(1-x)P_{111}^{00}(m_{1},m_{2},m_{3})  \nonumber \\
&&+P_{111}^{01}(m_{1},m_{2},m_{3}) +\frac{1}{2}%
T_{1}(m_{1}^{2})T_{1}(m_{2}^{2})-\frac{1}{2}T_{1}(m_{2}^{2})T_{1}(m_{3}^{2})+%
\frac{1}{2}T_{1}(m_{1}^{2})T_{1}(m_{3}^{2})  \nonumber \\
&&+\frac{1}{2}(m_{1}^{2}-m_{2}^{2}-\ell ^{\prime
2}-m_{3}^{2})P_{111}^{00}(m_{1},m_{2},m_{3}) -P_{111}^{10}(m_{1},m_{2},m_{3})
\nonumber \\
&&-P_{111}^{01}(m_{1},m_{2},m_{3}) -(1-x)(p_{M}\cdot
k)T_{2}(m_{1}^{2})T_{1}(m_{3}^{2}) -(m_{1}^{2}+m_{2}^{2}-\ell ^{\prime
2}-m_{3}^{2})\times  \nonumber \\
&&[\frac{(p_{M}\cdot \ell ^{\prime })}{\ell ^{\prime 2}}%
P_{211}^{10}(m_{1},m_{2},m_{3})-(p_{M}\cdot
k)(1-x)P_{211}^{00}(m_{1},m_{2},m_{3})  \nonumber \\
&&+P_{211}^{01}(m_{1},m_{2},m_{3}) +\frac{1}{2}%
T_{2}(m_{1}^{2})T_{1}(m_{2}^{2})-\frac{1}{2}P_{111}^{00}(m_{1},m_{2},m_{3})+%
\frac{1}{2}T_{2}(m_{1}^{2})T_{1}(m_{3}^{2})  \nonumber \\
&&+\frac{1}{2}(m_{1}^{2}-m_{2}^{2}-\ell ^{\prime
2}-m_{3}^{2})P_{211}^{00}(m_{1},m_{2},m_{3}) -P_{211}^{10}(m_{1},m_{2},m_{3})
\nonumber \\
&& -P_{211}^{01}(m_{1},m_{2},m_{3})] +2\frac{(p_{M}\cdot \ell ^{\prime })}{%
\ell ^{\prime 2}}P_{211}^{20}(m_{1},m_{2},m_{3})  \nonumber \\
&&-2(p_{M}\cdot
k)(1-x)P_{211}^{10}(m_{1},m_{2},m_{3})+2P_{211}^{11}(m_{1},m_{2},m_{3})
\nonumber \\
&&-P_{111}^{10}(m_{1},m_{2},m_{3})+(m_{1}^{2}-m_{2}^{2}-\ell ^{\prime
2}-m_{3}^{2})P_{211}^{10}(m_{1},m_{2},m_{3})  \nonumber \\
&&-2P_{211}^{20}(m_{1},m_{2},m_{3}) -2P_{211}^{11}(m_{1},m_{2},m_{3})
\nonumber \\
&&+2\frac{(p_{M}\cdot \ell ^{\prime })}{\ell ^{\prime 2}}%
P_{211}^{11}(m_{1},m_{2},m_{3})-2(p_{M}\cdot
k)(1-x)P_{211}^{01}(m_{1},m_{2},m_{3})  \nonumber \\
&&+2P_{211}^{02}(m_{1},m_{2},m_{3}) -P_{111}^{01}(m_{1},m_{2},m_{3})-\ell
^{\prime 2}T_{2}(m_{1}^{2})T_{1}(m_{3}^{2})  \nonumber \\
&&+(m_{1}^{2}-m_{2}^{2}-\ell ^{\prime
2}-m_{3}^{2})P_{211}^{01}(m_{1},m_{2},m_{3})
-2P_{211}^{11}(m_{1},m_{2},m_{3})-2P_{211}^{02}(m_{1},m_{2},m_{3})  \nonumber
\\
&&+\frac{2(p_{M}\cdot \ell ^{\prime })}{\ell ^{\prime 4}}%
P_{211}^{21}(m_{1},m_{2},m_{3})-2\frac{(p_{M}\cdot k)(1-x)}{\ell ^{\prime 2}}%
P_{211}^{11}(m_{1},m_{2},m_{3})  \nonumber \\
&&+\frac{2}{\ell ^{\prime 2}}P_{211}^{12}(m_{1},m_{2},m_{3}) -\frac{1}{\ell
^{\prime 2}}P_{111}^{11}(m_{1},m_{2},m_{3})  \nonumber \\
&&+\frac{(m_{1}^{2}-m_{2}^{2}-\ell ^{\prime 2}-m_{3}^{2})}{\ell ^{\prime 2}}%
P_{211}^{11}(m_{1},m_{2},m_{3})-\frac{2}{\ell ^{\prime 2}}%
P_{211}^{21}(m_{1},m_{2},m_{3})  \nonumber \\
&&\left.-\frac{2}{\ell ^{\prime
2}}P_{211}^{12}(m_{1},m_{2},m_{3})-\frac{1}{\ell
^{\prime 2}}T_{1}(m_{3}^{2})\int \frac{d^{n}p(p\cdot \ell ^{\prime })^{2}}{%
(p^{2}+m_{1}^{2})^{2}}\right\}+\cdots\,,
\end{eqnarray}
where $\ell ^{\prime }=\ell +kx$, $n=4-2\epsilon $,  $P_{M}$
represents the meson momentum, and $\{\cdots\}$ correspond to the
terms without $g^{\mu\nu}$.

\newpage

\section*{Appendix C}

\bigskip 
In this Appendix, we will 
replace the set of ten functions
$P_{211}^{ab}(m_{1},m_{2},m_{3};\ell ^{2})$ by the following
equivalent one of $H_{i}(m_{1},m_{2},m_{3};\ell ^{2})$ \cite {GY}.
The functions $H_{i}$ are free of quadratic divergencies and for
this reason they have simpler integral representations.

For the well known one-loop integrals with the definition $\gamma
_{s}=\gamma -1-\ln (4\pi) $, we have


\begin{eqnarray}
I(m^{2})\equiv \mu ^{4-D}\int \frac{d^{D}q}{(2\pi )^{D}}\frac{i}{q^{2}-m^{2}}
\hspace{9cm}  \nonumber \\
=\frac{m^{2}}{16\pi ^{2}}\left(\frac{m^{2}}{4\pi \mu ^{2}}\right)^{-\epsilon
}\Gamma (-1+\epsilon )\hspace{9.6cm}  \nonumber \\
=\frac{-m^{2}}{16\pi ^{2}}\left\{\frac{1}{\varepsilon }-\gamma _{s}-\ln
\left(\frac{m^{2}}{\mu ^{2}}\right)+\epsilon \left[ \frac{\pi ^{2}}{12}%
-\gamma _{s}-\ln \left(\frac{m^{2}}{\mu
^{2}}\right)+\frac{1}{2}\left(\gamma _{s} +1+\ln
\left(\frac{m^{2}}{\mu ^{2}}\right)\right)^{2}\right]\right\},%
\label{Im}
\end{eqnarray}

\begin{eqnarray}
(2\pi )^{4-n}T_{1}(m^{2})=(2\pi \mu )^{4-n}\int d^{n}p\frac{1}{p^{2}+m^{2}}%
\hspace{7.2cm}  \nonumber \\
=-m^{2}\pi ^{2}\left\{\frac{1}{\epsilon }-\gamma _{s}-\ln \left(\frac{m^{2}}{%
\mu ^{2}}\right)+\epsilon \left[ \frac{\pi ^{2}}{12}-\gamma _{s}-\ln \left(%
\frac{m^{2}}{\mu ^{2}}\right)+\frac{1}{2}\left(\gamma _{s}+1+\ln \left(\frac{%
m^{2}}{\mu ^{2}}\right)\right)^{2}\right]\right\},%
\label{Tm1}
\end{eqnarray}
\begin{eqnarray}
(2\pi )^{4-n}T_{2}(m^{2})=(2\pi \mu )^{4-n}\int d^{n}p\frac{1}{%
(p^{2}+m^{2})^{2}}\hspace{6.5cm}  \nonumber \\
=\pi ^{2}\left\{\frac{1}{\epsilon }-1-\gamma _{s}-\ln \left(\frac{m^{2}}{\mu
^{2}}\right)+\epsilon \left[ \frac{\pi ^{2}}{12}+\frac{1}{2}+\gamma _{s}+%
\frac{\gamma _{s}^{2}}{2}+\ln \left(\frac{m^{2}}{\mu
^{2}}\right)+\gamma _{s} \ln \left(\frac{m^{2}}{\mu
^{2}}\right)\right.\right.  \nonumber \\ \left.\left.
+\frac{1}{2}\ln \left(\frac{m^{2}}{\mu ^{2}}\right)\times \ln
\left(\frac{m^{2}}{\mu ^{2}}\right)\right]\right\}\,. \hspace{7cm}%
\label{Tm2}
\end{eqnarray}
We note that the expressions of Eq. (\ref{Im}) and Eqs.
(\ref{Tm1}) and (\ref{Tm2}) are defined in Minkowski and Euclidian
spaces, respectively.

For the two-point functions, the relations between $P_{111}^{ab}$
and $H_{i}$ are given by

\begin{eqnarray}
P_{111}^{00}(m_{1},m_{2},m_{3};\ell
^{2})&=&\frac{-1}{n-3}\{(m_{1}^{2}+\ell
^{2})H_{1}(m_{1},m_{2},m_{3})+H_{2}(m_{1},m_{2},m_{3})  \nonumber
\\
&&+m_{2}^{2}H_{1}(m_{2},m_{1},m_{3})+m_{3}^{2}H_{1}(m_{3},m_{1},m_{2})\}\,,
\\ \nonumber \\ P_{111}^{10}(m_{1},m_{2},m_{3};\ell
^{2})&=&P_{111}^{01}(m_{2},m_{1},m_{3};\ell ^{2})  \nonumber \\
&=& \frac{-1}{n-\frac{5}{2}}\left[\frac{\ell ^{2}}{2}%
P_{111}^{00}(m_{1},m_{2},m_{3})-P_{211}^{20}(m_{1},m_{2},m_{3})\right.
\nonumber \\ &&-m_{1}^{2}H_{2}(m_{1},m_{2},m_{3}) -m_{1}^{2}\ell
^{2}H_{1}(m_{1},m_{2},m_{3})  \nonumber \\
&&-m_{2}^{2}H_{3}(m_{2},m_{1},m_{3})
+m_{3}^{2}H_{2}(m_{3},m_{2},m_{1}) \nonumber \\
&&\left.+m_{3}^{2}H_{3}(m_{3},m_{2},m_{1})\right]\,, \\ \nonumber
\\
P_{111}^{11}(m_{1},m_{2},m_{3};\ell ^{2})&=&\frac{-1}{n-2}\left[%
m_{2}^{2}P_{211}^{11}(m_{2},m_{1},m_{3})-m_{3}^{2}P_{211}^{11}(m_{3},m_{2},m_{1})
\right. \nonumber \\
&&-m_{3}^{2}P_{211}^{02}(m_{3},m_{2},m_{1})-\ell
^{2}m_{3}^{2}P_{211}^{01}(m_{3},m_{2},m_{1})  \nonumber \\ &&
+m_{1}^{2}P_{211}^{11}(m_{1},m_{2},m_{3})
-P_{211}^{21}(m_{1},m_{2},m_{3}) \nonumber \\ &&\left.+\frac{\ell
^{2}}{2}P_{111}^{01}(m_{1},m_{2},m_{3})\right]\,, \\ \nonumber \\
P_{111}^{02}(m_{1},m_{2},m_{3};\ell
^{2})&=&P_{111}^{20}(m_{2},m_{1},m_{3};\ell ^{2})  \nonumber \\
&=&\frac{-1}{n-2}\left[%
m_{2}^{2}P_{211}^{20}(m_{2},m_{1},m_{3})+m_{3}^{2}P_{211}^{02}(m_{3},m_{2},m_{1}) \right.
\nonumber \\
&&%
\left.+m_{1}^{2}P_{211}^{02}(m_{1},m_{2},m_{3})-P_{211}^{12}(m_{1},m_{2},m_{3})
\right]\,.
\end{eqnarray}
For the three-point functions, we use \cite{GY}

\begin{eqnarray}
P_{211}^{00}(m_{1},m_{2},m_{3};\ell
^{2})&=&H_{1}(m_{1},m_{2},m_{3}) \,,\\ \nonumber \\
P_{211}^{10}(m_{1},m_{2},m_{3};\ell
^{2})&=&-H_{2}(m_{1},m_{2},m_{3})-\ell
^{2}H_{1}(m_{1},m_{2},m_{3})\,, \\ \nonumber \\
P_{211}^{01}(m_{1},m_{2},m_{3};\ell
^{2})&=&-H_{3}(m_{1},m_{2},m_{3})\,, \\ \nonumber \\
P_{211}^{20}(m_{1},m_{2},m_{3};\ell ^{2})&=&H_{4}(m_{1},m_{2},m_{3})+\frac{%
\ell ^{2}}{n}\left\{\left[(n-1)\ell
^{2}-m_{1}^{2}\right]H_{1}(m_{1},m_{2},m_{3})\right. \nonumber
\\%
&&\left.+2(n-1)H_{2}(m_{1},m_{2},m_{3})+P_{111}^{00}(m_{1},m_{2},m_{3})\right\}\,,
\\ \nonumber \\%
 P_{211}^{11}(m_{1},m_{2},m_{3};\ell
^{2})&=&H_{5}(m_{1},m_{2},m_{3})+\ell ^{2}H_{3}(m_{1},m_{2},m_{3})
\nonumber \\ &&+\frac{\ell
^{2}}{2n}\left[(m_{1}^{2}+m_{2}^{2}-m_{3}^{2}+\ell
^{2})H_{1}(m_{1},m_{2},m_{3})\right.  \nonumber \\
&&+2H_{2}(m_{1},m_{2},m_{3})-P_{111}^{00}(m_{1},m_{2},m_{3})
\nonumber \\
&&\left.+T_{2}(m_{1}^{2})T_{1}(m_{2}^{2})-T_{2}(m_{1}^{2})T_{1}(m_{3}^{2})\right\}\,,
\\ \nonumber \\
P_{211}^{02}(m_{1},m_{2},m_{3};\ell ^{2})&=&H_{6}(m_{1},m_{2},m_{3})+\frac{%
\ell ^{2}}{n}\left[-m_{2}^{2}H_{1}(m_{1},m_{2},m_{3})\right.
\nonumber
\\ && \left.+T_{2}(m_{1}^{2})T_{1}(m_{3}^{2})\right]\,, \\ \nonumber \\
P_{211}^{30}(m_{1},m_{2},m_{3};\ell
^{2})&=&-H_{7}(m_{1},m_{2},m_{3}) -P_{111}^{10}(m_{1},m_{2},m_{3})
\nonumber \\ &&-\frac{3\ell
^{2}}{n+2}\left\{\left(\frac{n-1}{3}\ell ^{2}-m_{1}^{2}\right)\ell
^{2}H_{1}(m_{1},m_{2},m_{3})\right.  \nonumber \\%
 &&+\left.\left[(n-1)\ell
^{2}-m_{1}^{2}\right]H_{2}(m_{1},m_{2},m_{3})+nH_{4}(m_{1},m_{2},m_{3})
\right\}\,, \\ \nonumber \\
P_{211}^{21}(m_{1},m_{2},m_{3};\ell ^{2})&=&-H_{8}(m_{1},m_{2},m_{3})-\frac{%
3\ell ^{2}}{n+2}\left[\frac{2}{3}(n-1)H_{5}(m_{1},m_{2},m_{3})
\right.\nonumber \\ &&+\left.\left(\frac{n-1}{3}\ell
^{2}-m_{1}^{2}\right)H_{3}(m_{1},m_{2},m_{3})-P_{111}^{01}(m_{1},m_{2},m_{3})\right]
\nonumber \\ &&-\frac{n-1}{n(n+2)}\ell
^{4}\left[(m_{1}^{2}+m_{2}^{2}-m_{3}^{2}+\ell
^{2})H_{1}(m_{1},m_{2},m_{3}) \right. \nonumber \\
&&+2H_{2}(m_{1},m_{2},m_{3})-P_{111}^{00}(m_{1},m_{2},m_{3})
\nonumber \\ &&\left.+T_{2}(m_{1}^{2})T_{1}(m_{2}^{2})
-T_{2}(m_{1}^{2})T_{1}(m_{3}^{2})\right]\,, \\ \nonumber \\
P_{211}^{12}(m_{1},m_{2},m_{3};\ell
^{2})&=&-H_{9}(m_{1},m_{2},m_{3})-\ell
^{2}H_{6}(m_{1},m_{2},m_{3})  \nonumber \\
&&-\frac{\ell ^{2}}{n+2}\left[2H_{5}(m_{1},m_{2},m_{3}) +\left(\frac{2\ell ^{2}}{n}%
-m_{2}^{2}\right)H_{2}(m_{1},m_{2},m_{3})\right.  \nonumber \\
&&+(m_{1}^{2}+m_{2}^{2}-m_{3}^{2}+\ell
^{2})H_{3}(m_{1},m_{2},m_{3}) \nonumber \\
&&\left. +P_{111}^{01}(m_{1},m_{2},m_{3})\right] -\frac{\ell ^{4}}{n(n+2)}%
\left\{-P_{111}^{00}(m_{1},m_{2},m_{3})\right.  \nonumber \\ &&
\left[m_{1}^{2}-(n+1)m_{2}^{2}-m_{3}^{2}+\ell
^{2}\right]H_{1}(m_{1},m_{2},m_{3}) \nonumber \\
&&\left.+T_{2}(m_{1}^{2})T_{1}(m_{2}^{2})
-(n-1)T_{2}(m_{1}^{2})T_{1}(m_{3}^{2})\right\}\,,
\\
\nonumber \\
P_{211}^{03}(m_{1},m_{2},m_{3};\ell ^{2})&=&-H_{10}(m_{1},m_{2},m_{3})-\frac{%
3\ell ^{2}}{n+2}\left[-m_{2}^{2}H_{3}(m_{1},m_{2},m_{3})
\right.\nonumber
\\ &&\left. +\ell ^{2}T_{2}(m_{1}^{2})T_{1}(m_{3}^{2})\right]\,.
\end{eqnarray}
The functions of $H_i$ are expressed as follows

\begin{eqnarray}
H_{1}(m_{1},m_{2},m_{3};\ell ^{2}) &=&\pi ^{4}\left[ \frac{2}{\Delta ^{2}}-%
\frac{1}{\Delta }(1-2\gamma _{m_{1}})-\frac{1}{2}+\frac{\pi
^{2}}{12}-\gamma _{m_{1}}+\gamma _{m_{1}}^{2}\right.  \nonumber \\
&&~~~~\left. +h_{1}(m_{1},m_{2},m_{3})\right] \,,\\ &&  \nonumber
\\
H_{2}(m_{1},m_{2},m_{3};\ell ^{2}) &=&\pi ^{4}\ell ^{2}\left[ -\frac{2}{%
\Delta ^{2}}+{\frac{1}{\Delta }}({\frac{1}{2}}-2\gamma _{m_{1}})+\frac{13}{8}%
-\frac{\pi ^{2}}{12}+{\frac{\gamma _{m_{1}}}{2}}-\gamma
_{m_{1}}^{2}\right. \nonumber \\ &&~~~~~~\left.
-h_{2}(m_{1},m_{2},m_{3})\right] \,,\\ &&  \nonumber \\
H_{3}(m_{1},m_{2},m_{3};\ell ^{2}) &=&\pi ^{4}\ell ^{2}\left[ \frac{1}{%
\Delta ^{2}}-\frac{1}{\Delta }(\frac{1}{4}-\gamma _{m_{1}})-\frac{13}{16}+%
\frac{\pi ^{2}}{24}-\frac{\gamma _{m_{1}}}{4}+\frac{\gamma _{m_{1}}^{2}}{2}%
\right.  \nonumber \\ &&\left.
~~~~~~+h_{3}(m_{1},m_{2},m_{3})\right]\,, \\ &&  \nonumber \\
H_{4}(m_{1},m_{2},m_{3};\ell ^{2}) &=&\pi ^{4}\ell ^{4}\left[ \frac{3}{%
2\Delta ^{2}}+\frac{1}{\Delta }\frac{3\gamma _{m_{1}}}{2}-\frac{175}{96}+%
\frac{\pi ^{2}}{16}+\frac{3\gamma _{m_{1}}^{2}}{4}\right.
\nonumber \\ &&\left.
~~~~~~+\frac{3}{4}h_{4}(m_{1},m_{2},m_{3})\right]\,, \\ &&
\nonumber \\
H_{5}(m_{1},m_{2},m_{3};\ell ^{2}) &=&\pi ^{4}\ell ^{4}\left[ -\frac{3}{%
4\Delta ^{2}}-\frac{1}{\Delta }\frac{3\gamma _{m_{1}}}{4}+\frac{175}{192}-%
\frac{\pi ^{2}}{32}-\frac{3\gamma _{m_{1}}^{2}}{8}\right.
\nonumber \\ &&\left.
~~~~~~-\frac{3}{4}h_{5}(m_{1},m_{2},m_{3})\right]\,, \\ &&
\nonumber \\
H_{6}(m_{1},m_{2},m_{3};\ell ^{2}) &=&\pi ^{4}\ell ^{4}\left[ \frac{1}{%
2\Delta ^{2}}-\frac{1}{\Delta }(\frac{1}{24}-\frac{\gamma _{m_{1}}}{2})-%
\frac{19}{32}+\frac{\pi ^{2}}{48}-\frac{\gamma _{m_{1}}}{24}\right.
\nonumber \\
&&\left. ~~~~~~+\frac{\gamma _{m_{1}}^{2}}{4}+\frac{3}{4}%
h_{6}(m_{1},m_{2},m_{3})\right] \,,\\ &&  \nonumber \\
H_{7}(m_{1},m_{2},m_{3};\ell ^{2}) &=&\pi ^{4}\ell ^{6}\left[ -\frac{1}{%
\Delta ^{2}}-\frac{1}{\Delta }(\frac{5}{24}+\gamma _{m_{1}})+\frac{287}{192}-%
\frac{\pi ^{2}}{24}-\frac{5\gamma _{m_{1}}}{24}\right.  \nonumber \\
&&\left. ~~~~~~-\frac{\gamma _{m_{1}}^{2}}{2}-\frac{1}{2}%
h_{7}(m_{1},m_{2},m_{3})\right] \,,\\ &&  \nonumber \\
H_{8}(m_{1},m_{2},m_{3};\ell ^{2}) &=&\pi ^{4}\ell ^{6}\left[ \frac{1}{%
2\Delta ^{2}}+\frac{1}{\Delta }(\frac{5}{48}+\frac{\gamma _{m_{1}}}{2})-%
\frac{287}{384}+\frac{\pi ^{2}}{48}+\frac{5\gamma _{m_{1}}}{48}\right.
\nonumber \\
&&\left. ~~~~~~+\frac{\gamma _{m_{1}}^{2}}{4}+\frac{1}{2}%
h_{8}(m_{1},m_{2},m_{3})\right] \,,\\ &&  \nonumber \\
H_{9}(m_{1},m_{2},m_{3};\ell ^{2}) &=&\pi ^{4}\ell ^{6}\left[ -\frac{1}{%
3\Delta ^{2}}-\frac{1}{\Delta }(\frac{1}{24}+\frac{\gamma _{m_{1}}}{3})+%
\frac{95}{192}-\frac{\pi ^{2}}{72}-\frac{\gamma _{m_{1}}}{24}\right.
\nonumber \\
&&\left. ~~~~~~-\frac{\gamma _{m_{1}}^{2}}{6}-\frac{1}{2}%
h_{9}(m_{1},m_{2},m_{3})\right] \,,\\ &&  \nonumber \\
H_{10}(m_{1},m_{2},m_{3};\ell ^{2}) &=&\pi ^{4}\ell ^{6}\left[ \frac{1}{%
4\Delta ^{2}}+\frac{1}{\Delta }(\frac{1}{96}+\frac{\gamma _{m_{1}}}{4})-%
\frac{283}{768}+\frac{\pi ^{2}}{96}+\frac{\gamma _{m_{1}}}{96}\right.
\nonumber \\
&&\left. ~~~~~~+\frac{\gamma _{m_{1}}^{2}}{8}+\frac{1}{2}%
h_{10}(m_{1},m_{2},m_{3})\right]\,,
\end{eqnarray}
where $\Delta =-2\epsilon $ and $\gamma _{m}=\gamma +\ln (\pi
m^{2}/\mu ^{2})$.

The ultraviolet finite parts $h_{i}(m_{1},m_{2},m_{3})$ of the function $%
H_{i}(m_{1},m_{2},m_{3};\ell ^{2})$ have the following one-dimensional
integral representations:
\begin{eqnarray}
h_{1}(m_{1},m_{2},m_{3})&=&\int_{0}^{1}dy[g(y)] \,, \nonumber \\
\nonumber \\
h_{2}(m_{1},m_{2},m_{3})&=&\int_{0}^{1}dy[g(y)+f_{1}(y)]\,,
\nonumber
\\ \nonumber \\
h_{3}(m_{1},m_{2},m_{3})&=&\int_{0}^{1}dy[g(y)+f_{1}(y)](1-y)\,,
\nonumber \\ \nonumber \\
h_{4}(m_{1},m_{2},m_{3})&=&\int_{0}^{1}dy[g(y)+f_{1}(y)+f_{2}(y)]\,,
\nonumber
\\
\nonumber \\
h_{5}(m_{1},m_{2},m_{3})&=&\int_{0}^{1}dy[g(y)+f_{1}(y)+f_{2}(y)](1-y)\,,
\nonumber \\ \nonumber \\
h_{6}(m_{1},m_{2},m_{3})&=&\int_{0}^{1}dy[g(y)+f_{1}(y)+f_{2}(y)](1-y)^{2}\,,
\nonumber \\ \nonumber \\
h_{7}(m_{1},m_{2},m_{3})&=&\int_{0}^{1}dy[g(y)+f_{1}(y)+f_{2}(y)+f_{3}(y)]\,,
\nonumber \\ \nonumber \\
h_{8}(m_{1},m_{2},m_{3})&=&%
\int_{0}^{1}dy[g(y)+f_{1}(y)+f_{2}(y)+f_{3}(y)](1-y) \,, \nonumber
\\ \nonumber \\
h_{9}(m_{1},m_{2},m_{3})&=&%
\int_{0}^{1}dy[g(y)+f_{1}(y)+f_{2}(y)+f_{3}(y)](1-y)^{2}\,,
\nonumber \\ \nonumber \\
h_{10}(m_{1},m_{2},m_{3})&=&%
\int_{0}^{1}dy[g(y)+f_{1}(y)+f_{2}(y)+f_{3}(y)](1-y)^{3}\,.
\end{eqnarray}
All ten integral representations are built up by the following
four basic functions:

\begin{eqnarray}
g(y)&=&Sp\left(\frac{1}{1-y_{1}}\right)+Sp\left(\frac{1}{1-y_{2}}\right)
+y_{1}\ln \left(\frac{y_{1}}{y_{1}-1}\right)+y_{2}\ln \left(\frac{y_{2}}{%
y_{2}-1}\right)\,,  \nonumber \\
\nonumber \\
f_{1}(y)&=&\frac{1}{2}\left[-\frac{1-\nu ^{2}}{\kappa ^{2}}+y_{1}^{2}\ln
\left(\frac{y_{1}}{y_{1}-1}\right)+y_{2}^{2}\ln \left(\frac{y_{2}}{y_{2}-1}%
\right)\right]\,,  \nonumber \\
\nonumber \\
f_{2}(y)&=&\frac{1}{3}\left[-\frac{2}{\kappa ^{2}}-\frac{1-\nu ^{2}}{2\kappa
^{2}}-\left(\frac{1-\nu ^{2}}{\kappa ^{2}}\right)^{2}+y_{1}^{3}\ln \left(%
\frac{y_{1}}{y_{1}-1}\right)+y_{2}^{3}\ln \left(\frac{y_{2}}{y_{2}-1}\right)%
\right]\,,  \nonumber \\
\nonumber \\
f_{3}(y)&=&\frac{1}{4}\left[-\frac{4}{\kappa ^{2}}-\left(\frac{1}{3}+\frac{3%
}{\kappa ^{2}}\right)\frac{1-\nu ^{2}}{2\kappa ^{2}}-\frac{1}{2}\left(\frac{%
1-\nu ^{2}}{\kappa ^{2}}\right)^{2}-\left(\frac{1-\nu ^{2}}{\kappa ^{2}}%
\right)^{3}\right.  \nonumber \\
&&\left.~~~~+y_{1}^{4}\ln \left(\frac{y_{1}}{y_{1}-1}\right)+y_{2}^{4}\ln \left(%
\frac{y_{2}}{y_{2}-1}\right)\right]\,,
\end{eqnarray}
where
\begin{eqnarray}
Sp(z)&=&\int_{0}^{z}-\frac{\ln
(1-t)}{t}dt\,,\nonumber \\
\nonumber \\
y_{1,2}&=&\frac{1+\kappa ^{2}-\nu ^{2}\pm
\sqrt{(1+\kappa ^{2}-\nu ^{2})^{2}+4\nu ^{2}\kappa ^{2}-4i\kappa
^{2}\eta }}{2\kappa ^{2}}\,,
\nonumber \\
\nonumber \\
\nu ^{2}&=&\frac{ay+b(1-y)}{y(1-y)}\,,\ \ a=\frac{m_{2}^{2}}{m_{1}^{2}}\,,\
\ b=\frac{m_{3}^{2}}{m_{1}^{2}}\,,\ \ \kappa ^{2}=\frac{\ell ^{2}}{m_{1}^{2}}%
\,.
\end{eqnarray}

Finally, we must transform the parameters back into the Minkowski
space and change the inward directions of the momenta $l,k$ for
the final particles by outward ones:

\begin{eqnarray}
\ell ^{2}&\rightarrow& -(m_{K,\pi }^{2}-2p\cdot k)\,,  \nonumber \\
\nonumber \\
\ell ^{\prime 2}&\rightarrow &-[m_{K,\pi }^{2}-2p\cdot k(1-x)]\,,  \nonumber
\\
\nonumber \\ p\cdot k&\rightarrow &p\cdot k\,,  \nonumber \\
\nonumber \\ p\cdot \ell ^{\prime }&\rightarrow &m_{K,\pi
}^{2}-p\cdot k(1-x)\,, \nonumber \\ \nonumber \\ k\cdot \ell
^{\prime }&\rightarrow& -p\cdot k\,,
\end{eqnarray}
where $0\leq p\cdot k\leq (m_{K,\pi }^{2}-m_{\ell }^{2})/2$, and
we have replaced $\partial ^{2}$ by $-\partial ^{2}$ as calculated
in the Euclidian space.

\end{document}